\documentstyle[epsf]{mn}

\newif\ifAMStwofonts
\newcommand{\hi}{\hbox{H\,{\sc i}}}
\newcommand{\hii}{\hbox{H\,{\sc ii}}}
\newcommand{\hei}{\hbox{He\,{\sc i}}}
\newcommand{\heii}{\hbox{He\,{\sc ii}}}
\newcommand{\oh}{\hbox{O/H}}

\newcommand{\ha}{\hbox{H$\alpha$}}
\newcommand{\hanii}{\hbox{H$\alpha$+[N\,{\sc ii}]}}
\newcommand{\hb}{\hbox{H$\beta$}}
\newcommand{\hg}{\hbox{H$\gamma$}}
\newcommand{\hd}{\hbox{H$\delta$}}
\newcommand{\oi}{\hbox{[O\,{\sc i}]}}
\newcommand{\oii}{\hbox{[O\,{\sc ii}]}}
\newcommand{\oiii}{\hbox{[O\,{\sc iii}]}}
\newcommand{\nii}{\hbox{[N\,{\sc ii}]}}
\newcommand{\sii}{\hbox{[S\,{\sc ii}]}}
\newcommand{\siii}{\hbox{[S\,{\sc iii}]}}
\newcommand{\nev}{\hbox{[Ne\,{\sc v}]}}
\newcommand{\fevii}{\hbox{[Fe\,{\sc vii}]}}
\newcommand{\mgii}{\hbox{Mg\,{\sc ii}}}
\newcommand{\feii}{\hbox{[Fe\,{\sc ii}]}}
\newcommand{\hahb}{\hbox{H$\alpha$/H$\beta$}}
\newcommand{\oiha}{\hbox{[O\,{\sc i}]/H$\alpha$}}
\newcommand{\oiiha}{\hbox{[O\,{\sc ii}]/H$\alpha$}}

\newcommand{\oiiihb}{\hbox{[O\,{\sc iii}]/H$\beta$}}
\newcommand{\oiioiii}{\hbox{[O\,{\sc ii}]/[O\,{\sc iii}]}}
\newcommand{\siiha}{\hbox{[S\,{\sc ii}]/H$\alpha$}}
\newcommand{\niiha}{\hbox{[N\,{\sc ii}]/H$\alpha$}}
\newcommand{\niisii}{\hbox{[N\,{\sc ii}]/[S\,{\sc ii}]}}

\newcommand{\etaha}{\hbox{$\eta_{{\rm H}\alpha}$}}
\newcommand{\etahb}{\hbox{$\eta_{{\rm H}\beta}$}}
\newcommand{\etauv}{\hbox{$e^{-\hat{\tau}_V}$}}

\newcommand{\etahanii}{\hbox{$\eta_{{\rm H}\alpha+[N\,{\sc ii}]}$}}
\newcommand{\oetahanii}{\hbox{$\eta_{{\rm H}\alpha+[N\,{\sc ii}]}^0$}}
\newcommand{\oetaha}{\hbox{$\eta_{{\rm H}\alpha}^0$}}
\newcommand{\etaoii}{\hbox{$\eta_{\rm [O\,{\sc ii}]}$}}
\newcommand{\oetaoii}{\hbox{$\eta_{\rm [O\,{\sc ii}]}^0$}}
\newcommand{\etaoiii}{\hbox{$\eta_{\rm [O\,{\sc iii}]}$}}
\newcommand{\dbal}{\hbox{D$_{4000}$}}
\newcommand{\wha}{\hbox{$W_{{\rm H}\alpha}$}}
\newcommand{\whanii}{\hbox{$W_{{\rm H}\alpha+[N\,{\sc ii}]}$}}
\newcommand{\whb}{\hbox{$W_{{\rm H}\beta}$}}
\newcommand{\woii}{\hbox{$W_{\rm [O\,{\sc ii}]}$}}

\newcommand{\lhahb}{\hbox{$L_{{\rm H}\alpha}/L_{{\rm H}\beta}$}}
\newcommand{\lhaniihb}{\hbox{$L_{{\rm H}\alpha+{\rm [N\,{\sc ii}]}}/L_{{\rm H}\beta}$}}
\newcommand{\loiiihb}{\hbox{$L_{\rm [O\,{\sc iii}]}/L_{{\rm H}\beta}$}}
\newcommand{\loiiha}{\hbox{$L_{\rm [O\,{\sc ii}]}/L_{{\rm H}\alpha}$}}
\newcommand{\loiihb}{\hbox{$L_{\rm [O\,{\sc ii}]}/L_{{\rm H}\beta}$}}
\newcommand{\loiihanii}{\hbox{$L_{\rm [O\,{\sc ii}]}/L_{{\rm H}\alpha+{\rm [N\,{\sc ii}]}}$}}
\newcommand{\loiiihanii}{\hbox{$L_{\rm [O\,{\sc iii}]}/L_{{\rm H}\alpha+{\rm [N\,{\sc ii}]}}$}}
\newcommand{\lniiha}{\hbox{$L_{\rm [N\,{\sc ii}]}/L_{{\rm H}\alpha}$}}
\newcommand{\loiioiii}{\hbox{$L_{\rm [O\,{\sc ii}]}/L_{\rm [O\,{\sc iii}]}$}}
\newcommand{\lsiiha}{\hbox{$L_{\rm [S\,{\sc ii}]}/L_{{\rm H}\alpha}$}}
\newcommand{\lsiihanii}{\hbox{$L_{\rm [S\,{\sc ii}]}/L_{{\rm H}\alpha+{\rm [N\,{\sc ii}]}}$}}
\newcommand{\lniisii}{\hbox{$L_{\rm [N\,{\sc ii}]}/L_{\rm [S\,{\sc ii}]}$}}
\newcommand{\lpoiiihb}{\hbox{$L_{\rm [O\,{\sc iii}]}^+/L_{{\rm H}\beta}^+$}}
\newcommand{\lpoiioiii}{\hbox{$L_{\rm [O\,{\sc ii}]}^+/L_{\rm [O\,{\sc iii}]}^+$}}
\newcommand{\lpsiiha}{\hbox{$L_{\rm [S\,{\sc ii}]}^+/L_{{\rm H}\alpha}^+$}}
\newcommand{\lpniisii}{\hbox{$L_{\rm [N\,{\sc ii}]}^+/L_{\rm [S\,{\sc ii}]}^+$}}
\newcommand{\uva}{\hbox{$\langle U\rangle$}}
\newcommand{\uav}{\hbox{$\langle \tilde{U}\rangle$}}
\newcommand{\uavo}{\hbox{$\langle {\tilde{U}}_0\rangle$}}
\newcommand{\qav}{\hbox{$\tilde{Q}$}}
\newcommand{\mav}{\hbox{${\tilde{M}}_\ast$}}
\newcommand{\eav}{\hbox{$\tilde{\epsilon}$}}
\newcommand{\nav}{\hbox{$\tilde{n}_{\rm H}$}}
\newcommand{\xav}{\hbox{$\tilde{\xi}_{\rm d}$}}
\newcommand{\zav}{\hbox{$\tilde{Z}$}}
\ifoldfss
  \ifCUPmtlplainloaded \else
    \NewTextAlphabet{textbfit} {cmbxti10} {}
    \NewTextAlphabet{textbfss} {cmssbx10} {}
    \NewMathAlphabet{mathbfit} {cmbxti10} {} % for math mode
    \NewMathAlphabet{mathbfss} {cmssbx10} {} %  "   "    "
  \fi
  \ifAMStwofonts
    \ifCUPmtlplainloaded \else
      \NewSymbolFont{upmath} {eurm10}
      \NewSymbolFont{AMSa} {msam10}
      \NewMathSymbol{\upi}     {0}{upmath}{19}
      \NewMathSymbol{\umu}     {0}{upmath}{16}
      \NewMathSymbol{\upartial}{0}{upmath}{40}
      \NewMathSymbol{\leqslant}{3}{AMSa}{36}
      \NewMathSymbol{\geqslant}{3}{AMSa}{3E}

      \let\leq=\leqslant \let\le=\leqslant
      \let\geq=\geqslant \let\ge=\geqslant
    \fi
  \fi
\fi % End of OFSS

\ifnfssone
  \newmathalphabet{\mathit}
  \addtoversion{normal}{\mathit}{cmr}{m}{it}
  \addtoversion{bold}{\mathit}{cmr}{bx}{it}
  \newmathalphabet{\mathbfit} % math mode version of \textbfit{..}
  \addtoversion{normal}{\mathbfit}{cmr}{bx}{it}
  \addtoversion{bold}{\mathbfit}{cmr}{bx}{it}
  \newmathalphabet{\mathbfss} % math mode version of \textbfss{..}
  \addtoversion{normal}{\mathbfss}{cmss}{bx}{n}
  \addtoversion{bold}{\mathbfss}{cmss}{bx}{n}
  \ifAMStwofonts
    \ifCUPmtlplainloaded \else
      \UseAMStwoboldmath
      \makeatletter
      \new@mathgroup\upmath@group
      \define@mathgroup\mv@normal\upmath@group{eur}{m}{n}
      \define@mathgroup\mv@bold\upmath@group{eur}{b}{n}
      \edef\UPM{\hexnumber\upmath@group}
      \new@mathgroup\amsa@group
      \define@mathgroup\mv@normal\amsa@group{msa}{m}{n}
      \define@mathgroup\mv@bold\amsa@group{msa}{m}{n}
      \edef\AMSa{\hexnumber\amsa@group}
      \makeatother
      \mathchardef\upi="0\UPM19
      \mathchardef\umu="0\UPM16
      \mathchardef\upartial="0\UPM40
      \mathchardef\leqslant="3\AMSa36
      \mathchardef\geqslant="3\AMSa3E

      \let\leq=\leqslant \let\le=\leqslant
      \let\geq=\geqslant \let\ge=\geqslant
    \fi
  \fi
\fi % End of NFSS release 1

\ifnfsstwo
  \DeclareMathAlphabet{\mathbfit}{OT1}{cmr}{bx}{it}
  \SetMathAlphabet\mathbfit{bold}{OT1}{cmr}{bx}{it}
  \DeclareMathAlphabet{\mathbfss}{OT1}{cmss}{bx}{n}
  \SetMathAlphabet\mathbfss{bold}{OT1}{cmss}{bx}{n}
  \ifAMStwofonts
    \ifCUPmtlplainloaded \else
      \DeclareSymbolFont{UPM}{U}{eur}{m}{n}
      \SetSymbolFont{UPM}{bold}{U}{eur}{b}{n}
      \DeclareSymbolFont{AMSa}{U}{msa}{m}{n}
      \DeclareMathSymbol{\upi}{0}{UPM}{"19}
      \DeclareMathSymbol{\umu}{0}{UPM}{"16}
      \DeclareMathSymbol{\upartial}{0}{UPM}{"40}
      \DeclareMathSymbol{\leqslant}{3}{AMSa}{"36}
      \DeclareMathSymbol{\geqslant}{3}{AMSa}{"3E}

      \let\leq=\leqslant \let\le=\leqslant
      \let\geq=\geqslant \let\ge=\geqslant
    \fi
  \fi
\fi % End of NFSS release 2

\ifCUPmtlplainloaded \else
  \ifAMStwofonts \else % If no AMS fonts
    \def\upi{\pi}
    \def\umu{\mu}
    \def\upartial{\partial}
  \fi
\fi

\title{Nebular Emission from Star-Forming Galaxies}
\author[S. Charlot \& M. Longhetti]
       {St\'ephane Charlot\thanks{Currently on leave at the Max-Planck Institut
	f\"ur Astrophysik, 
	Karl-Schwarzschild-Strasse 1, 85748 Garching, Germany} and 
	Marcella Longhetti\thanks{Present address: Osservatorio Astronomico di Brera,
        Via Bianchi 46, 23807 Merate (LC), Italy}\\
        Institut d'Astrophysique de Paris, CNRS, 98 bis Boulevard Arago, 75014 Paris, France\\
	}
\date{Accepted ... .
      Received ... ;
      in original form ...}

\pagerange{\pageref{firstpage}--\pageref{lastpage}}
\pubyear{1994}

\begin{document}

\maketitle

\label{firstpage}

\begin{abstract}
We present a new model for computing consistently the line and
continuum emission from galaxies, based on a combination of
recent population synthesis and photoionization codes.
We use effective parameters to describe the \hii\ regions and the 
diffuse gas ionized by single stellar generations in a galaxy, among
which the most important ones are the zero-age effective ionization
parameter, the effective gas metallicity, and the effective 
dust-to-heavy element ratio. We calibrate the nebular properties of
our model using the observed \oiiihb, \oiioiii, \siiha, and 
\niisii\ ratios of a representative sample of nearby spiral
and irregular, starburst, and \hii\ galaxies. To compute whole
(line plus continuum) spectral energy distributions, we include
the absorption by dust in the neutral interstellar medium (ISM)
using a recent simple prescription, which is consistent with 
observations of nearby starburst galaxies. Our model enables us
to interpret quantitatively the observed optical spectra of 
galaxies in terms of stars, gas, and dust parameters. We find
that the range of ionized-gas properties spanned by nearby 
galaxies implies factors of 3.5 and 14 variations in the \ha\
and \oii\ luminosities produced per unit star formation rate 
(SFR). When accounting for stellar \ha\ absorption and absorption
by dust in the neutral ISM, the actual uncertainties in SFR 
estimates based on the emergent \ha\ and \oii\ luminosities are as
high as several decades. We derive new estimators of the SFR, 
the gas-phase oxygen abundance, and the effective absorption
optical depth of the dust in galaxies. We show that, with the
help of other lines such as \oii, \hb, \oiii\, \nii, or \sii,
the uncertainties in SFR estimates based on \ha\ can be reduced
to a factor of only 2--3, even if the \ha\ line is blended with
the adjacent \nii\ lines.  Without \ha, however, the SFR is 
difficult to estimate from the \oii, \hb, and \oiii\ lines. The
reason for this is that the absorption by dust in the neutral ISM
and the ionized-gas parameters are then difficult to constrain 
independently. This suggests that, while insufficient by itself,
the \ha\ line is essential for estimating the star formation rate
from the optical emission of a galaxy.

\end{abstract}


\section{Introduction}

The most prominent tracers of star formation in the integrated
spectra of galaxies are the emission lines produced by short-lived,
massive stars. Several approximate formulae have been derived to 
convert the \ha\ and \oii\ luminosities of observed galaxies into
star formation rates (SFRs; see the review by Kennicutt 1998). 
Calibrations of the \ha\ luminosity emitted per unit SFR have 
consisted usually in applying dust-free case~B recombination to
the stellar ionizing radiation predicted by a population synthesis
model (e.g., Kennicutt 1983). The scaling of this luminosity by
some empirical \oiiha\ ratio provides a calibration of the \oii\
luminosity emitted per unit SFR (Gallagher, Hunter \& Bushouse 
1989; Kennicutt 1992b; see Barbaro \& Poggianti 1997 for a more 
refined calibration). In reality, the absorption of ionizing 
photons by dust in \hii\ regions could affect seriously the \ha\
luminosity emerging from a star-forming galaxy (e.g., Mathis 
1986b). Moreover, since the \oiiha\ ratio can vary by an order 
of magnitude in nearby star-forming galaxies, SFR estimates 
based on this ratio are subject to substantial uncertainties 
(Kennicutt 1998). To address these issues, we require a model 
for relating consistently the luminosities of emission lines to
the stars, gas, and dust parameters of galaxies.

The optical-line ratios in the integrated spectra of nearby 
star-forming galaxies are similar to those in the spectra of 
individual \hii\ regions (e.g., Lehnert \& Heckman 1994; 
Kobulnicky, Kennicutt \& Pizagno 1999). Photoionization models
have proved essential to interpret the emission-line properties
of \hii\ regions in terms of stars and gas parameters (e.g.,
Evans \& Dopita 1985). Thus, these models can potentially help
us also interpret the nebular emission from galaxies in terms
of macroscopic star formation parameters. In fact, 
photoionization models have already been combined with 
evolutionary population synthesis codes to constrain the ages
and metallicities of giant extragalactic \hii\ regions and 
\hii\ galaxies (e.g., Garc\'{\i}a-Vargas, Bressan \& D\'{\i}az
1995, Stasi\'nska \& Leitherer 1996, and references therein).
These studies, however, are not suited to spectral analyses of
galaxies containing several stellar generations. Furthermore,
until recently, combinations of photoionization and population
synthesis models were limited by the lack of a simple prescription
to compute consistently the effects of dust on the line and 
continuum emission from galaxies (see however Charlot \& Fall 
2000). It is now possible to build a model to interpret 
simultaneously the signatures of stars, gas, and dust in the
integrated spectra of galaxies. The purpose of this paper is to
present such a model.

We combine recent population synthesis and photoionization
codes to compute the line and continuum emission from galaxies.
We use effective parameters to describe the \hii\ regions
and the diffuse gas ionized by single stellar generations
in a galaxy. We first calibrate the nebular properties of
our model using the observed \oiiihb, \oiioiii, \siiha, and
\niisii\ ratios of a representative sample of nearby spiral
and irregular, starburst, and \hii\ galaxies. To compute
whole (line plus continuum) spectral energy distributions, we
then include the absorption by dust in the neutral interstellar
medium (ISM) using a simple prescription, which is consistent
with observations of nearby starburst galaxies (Charlot \& 
Fall 2000). Our model succeeds in accounting quantitatively 
for the optical line and continuum emission from nearby 
star-forming galaxies of various types. Such spectral fits 
provide insights simultaneously into the star formation history,
metallicity, and absorption by dust in the galaxies. This enables
us to identify and calibrate new estimators of the star formation
rate, the gas-phase oxygen abundance, and the effective absorption
optical depth of the dust in galaxies.

We present our model in \S2, where we express the nebular 
emission of a galaxy in terms of effective gas parameters.
In \S3, we compare our model with observations and identify
the specific influence of each parameter on the integrated 
spectral properties of galaxies. In \S4, we use the observed
relations between various integrated spectral properties of 
nearby star-forming galaxies to construct estimators of the 
star formation rate, the gas-phase oxygen abundance, and the
effective absorption optical depth of the dust. We express our
results in terms of simple formulae derived for different 
assumptions about the available spectral information. Our 
conclusions are summarized in \S5.

\section[]{The Model}
In this section, we describe our model for the emission from
stars and gas in galaxies. To begin with, in \S2.1--2.3, we
focus on the production of stellar radiation and its conversion
into emission lines and recombination continuum by the gas. 
We compute the luminosity emerging from a galaxy in \S2.4, where
we include the absorption of photons emanating from the ionized
regions by dust in the neutral (and molecular) ISM. We do not 
consider here the contribution to the ionizing radiation by active
galactic nuclei (AGNs; see \S4).

\subsection{Basic Assumptions}

The luminosity per unit wavelength $L_\lambda^+(t)$ produced
at time $t$ by the stars and the ionized gas in a galaxy can
be expressed generally as
\begin{equation}
L_\lambda^+(t)=\int_0^t\,dt'\,\psi(t-t')\,S_\lambda(t')
	\,T_\lambda^{+}(t,t').
\label{basic}
\end{equation}
Here $\psi(t-t')$ is the star formation rate at time $t-t'$, 
$S_\lambda(t')$ is the luminosity emitted per unit wavelength and 
per unit mass by a stellar generation of age $t'$, and $T_\lambda^{+}
(t,t')$ is the `transmission function' of the ionized gas, defined
as the fraction of the radiation produced at wavelength $\lambda$
at time $t$ by a generation of stars of age $t'$ that is transferred
by the gas these stars ionize. Thus, $T_\lambda^{+}(t,t')$ must be
regarded as the average transmission of the gas ionized by all stars
of age $t'$ within the galaxy. If the ionized regions are bounded by 
neutral material, $T_\lambda^{+} (t,t')$ will be close to zero 
at wavelengths blueward of the H-Lyman limit but greater than unity
at the wavelengths corresponding to emission lines. To compute 
$L_\lambda^+(t)$ from equation~(\ref{basic}), we make the following
simplifying assumptions.

We neglect the ionization of the gas by hot intermediate- and 
low-mass stars; we thus write
\begin{equation}
T_\lambda^{+}(t,t')=1, \hskip1truecm{\rm for}\,\,t'\geq t^{\rm OB}\,,
\label{tbp}
\end{equation}
where $t^{\rm OB}$ is the lifetime of massive stars producing most of the
ionizing photons (\S2.2). At ages greater than $t^{\rm OB}$, planetary
nebula nuclei are the dominant sources of ionizing photons in a single
stellar generation. The ionizing radiation from planetary nebula nuclei,
however, is typically less than 0.1 per cent of that produced at earlier
ages by massive stars, even for initial mass functions (IMFs) with steep
slopes and low upper cutoff masses (e.g., Charlot \& Fall 1993; Binette
et al. 1994). Since we are interested in galaxies in the process of 
forming massive stars, we therefore neglect the influence of the 
radiation from old planetary nebula nuclei on the gas.

We also neglect the influence of stellar winds and supernovae on the 
emission from galaxies. This is justified by both models and observations
of star-forming galaxies. Leitherer \& Heckman (1995) have computed the 
power injected into the ISM by stellar winds and supernovae for stellar
populations with various IMFs and star formation histories. For a normal
IMF, this amounts to less than 10 per cent of the power of the ionizing
radiation from massive stars (see Fig.~69 of Leitherer \& Heckman 1995).
Observations support these results. About 20 to 50 per cent of the 
H-Balmer line emission from nearby spiral and irregular galaxies arises
from `diffuse ionized gas', which is not directly associated to individual
\hii\ regions (e.g., Hunter \& Gallagher 1990; Hoopes, Walterbos \& 
Greenwalt 1996; Ferguson et al. 1996; Martin 1997). This gas, however,
appears to also be heated primarily by the ionizing radiation from massive 
stars rather than by shocks (Mathis 1986a; Hunter \& Gallagher 1990;
Hoopes et al. 1996; Martin 1997; Wang, Heckman \& Lehnert 1997). 
Direct evidence that 20 to 50 per cent of the ionizing photons produced 
in individual \hii\ regions might leak into the ambient ISM further 
supports this result (Oey \& Kennicutt 1997). Hence, for most purposes, 
the influence of stellar winds and supernovae on the nebular emission from
galaxies can probably be neglected.

Finally, we assume for simplicity that the transmission function of the
ionized gas depends only on the age $t'$ of the stars producing the 
ionizing radiation, i.e.
\begin{equation}
T_\lambda^{+}(t,t')=T_\lambda^{+}(t'), 
\hskip1truecm{\rm for}\,\,t'< t^{\rm OB}\,.
\label{ttp}
\end{equation}
This approximation is adequate for the purpose of interpreting
observations of star-forming galaxies. In fact, since the nebular 
emission depends only on the current gas content of the galaxies, 
specifying the dependence of $T_\lambda^{+}$ on the galaxy age
$t$ would involve arbitrary assumptions about the chemical enrichment
history. We now appeal to standard codes to compute the radiation
$S_\lambda(t')$ from the stars and the transmission function 
$T_\lambda^{+}(t')$ of the gas in our model.

\subsection{Stellar Radiation}

We compute the luminosity $S_\lambda(t')$ emitted per unit wavelength and
per unit mass by a stellar generation of age $t'$ using the most recent
version of the Bruzual \& Charlot (1993) population synthesis code. This
includes all phases of stellar evolution, from the zero-age main sequence 
to supernova explosions for progenitors more massive than $8\,M_{\odot}$,
and to the end of the white dwarf cooling sequence for less massive 
progenitors. Models can be computed at metallicities $Z=0.0001$, 0.0004,
0.004, 0.008, 0.02, 0.05, and 0.1 ($Z_\odot=0.02$). In the models used here, 
stars evolve along the Padova tracks (Alongi et al. 1993; Bressan et al. 
1993; Fagotto et al. 1994a,b,c; Girardi et al. 1996). The tracks include
mild overshooting in the convective cores of stars more massive than 
$1.0\, M_{\odot}$ (with a reduced efficiency between 1.0 and 1.5$\, M_\odot$),
as indicated by various observations of Galactic star clusters. We adopt
the semi-empirical library of stellar spectra compiled by Lejeune et al. 
(1997, 1998) to describe the emission from stars of all metallicities. This
library relies on model atmospheres by Kurucz (1992, 1995) for the hotter 
(O--K) stars with effective temperatures in the range $3,500<T_{\rm eff} 
\leq50,000$~K, complemented with models by Bessell et al. (1989, 1991) and
Fluks et al. (1994) for cooler M giants, and with models by Allard \& 
Hauschildt (1995) for M dwarfs. Lejeune et al. (1997, 1998) used empirical
corrections to refine all these models at optical and infrared wavelengths.
The spectra of stars hotter than $T_{\rm eff}=50,000\,$K in the population
synthesis models (i.e., a few short-lived Wolf-Rayet stars and central 
stars of planetary nebulae) are approximated by pseudo-blackbodies, 
which include the discontinuities across the important photoionization
edges blueward of the H-Lyman limit (as parameterized by Shields \& Searle
1978 to reproduce the models of Hummer \& Mihalas 1970). Nearly all the 
ionizing radiation from a stellar population with a normal IMF is produced 
by stars with $T_{\rm eff}\ga 30,000\,$K. Thus, the results presented in this
paper rely heavily on the Kurucz model atmospheres.

The Kurucz (1992, 1995) models are static, plane-parallel atmospheres in
local thermodynamic equilibrium (LTE) with line blanketing by over 58
million atomic and molecular transitions. They have been tested against
other recent models that include non-LTE effects, spherical geometry,
atmospheric expansion, and stellar winds, but which are generally
restricted to smaller ranges of temperature and include line blanketing
by fewer transitions (Vacca, Garmany \& Shull 1996, and references therein;
Schaerer \& de Koter 1997, and references therein). These comparisons
indicate that the Kurucz models produce more H-ionizing photons ($\lambda 
\leq 912\,${\AA}) than spherical models including non-LTE and wind effects,
by an amount that increases from a few percent at $T_{\rm eff}=50,000\,$K
to about 25 per cent at $T_{\rm eff}=35,000\,$K for main-sequence stars.
The Kurucz models produce also less \hei-ionizing photons ($\lambda \leq
504\,${\AA}), by an amount that decreases from 40 to 25 per cent in this 
temperature range. Thus, for a stellar population with a normal IMF, we 
expect the overall discrepancy in the H- and \hei-ionizing continua 
produced by the two types of models to be less than about 15 and 30 per
cent, respectively. This level of accuracy is sufficient for our purposes.
In fact, the absolute accuracy of the Kurucz models should be higher, since
the effects of line blanketing are probably underestimated in spherical
models including non-LTE and wind effects. Unfortunately, observations of
the H-ionizing continuum of stars with $T_{\rm eff}\ga 30,000\,$K are not
available, while the spectra of two stars with $T_{\rm eff}\ga 20,000\,$K 
have led to ambiguous interpretations (Cassinelli et al. 1995; Vacca et al. 
1996; Schaerer \& de~Koter 1997). We note that the neglect of wind effects
in the Kurucz models may result in a deficiency of \heii-ionizing photons
($\lambda \leq 228\, ${\AA}; see Schaerer \& de~Koter 1997). Therefore, 
we do not rely on the predictions of the population synthesis code at
wavelengths blueward of the \heii-ionizing edge.

The main adjustable parameters in the population synthesis code are
the IMF, the star formation rate, and the metallicity. For simplicity,
we assume that the IMF can be approximated by a single power
law, $\phi(m) \propto m^{-1-x}$ [defined such that $\phi(m)dm$ 
is the number of stars born with masses between $m$ and $m+dm$],
with an exponent $x$ and lower and upper cutoff masses $m_{\rm L}$
and $m_{\rm U}$. We include $m_{\rm U}$ as an adjustable parameter
and, unless otherwise indicated, adopt a Salpeter IMF ($x=1.35$) with
$m_{\rm L}=0.1\,M_\odot$. We compute a model spectrum at ages from 
$1\times 10^5$~yr to $2\times 10^{10}$~yr in time steps that increase
from $1.5 \times10^4$~yr to $2.5\times10^8$~yr at late ages. At each
time step, the spectrum is defined over the range of wavelengths from
91~{\AA} to 160~$\mu$m, with samplings of 10~{\AA} in the ultraviolet,
20~{\AA} in the visual, and increasing further in the infrared. The 
lifetimes of the stars producing most of the ionizing radiation are 
typically 3--5$\times 10^6\,$yr at all metallicities. To be conservative,
we adopt $t^{\rm OB} =1\times10^7\,$yr in equations~(\ref{tbp}) and 
(\ref{ttp}), corresponding to a drop by over 99 per cent in the 
production rate of ionizing photons. 

We are also interested in the absorption equivalent widths of the
low-order Balmer lines of \hi. The spectral resolution of the 
population synthesis code is too low for line strengths to be 
measured reliably. We adopt the standard procedure that consists
in parameterizing absorption-line strengths as functions of the
stellar effective temperature, gravity, and metallicity (e.g., 
Worthey et al. 1994). This method has been developed mainly to
interpret the features of late-type dwarf and giant stars in the
spectra of old stellar populations. As a consequence, appropriate
calibrations do not exist for the \ha\ and \hb\ absorption equivalent
widths of hot-main sequence and supergiant stars that dominate the
spectra of young stellar populations. Cananzi, Augarde \& Lequeux
(1993) did collect accurate \hg\ and \hd\ equivalent widths of nearby
stars over complete ranges of spectral types and luminosity classes
for the specific purpose of population synthesis analyses. Detailed 
Balmer-line models by Kurucz (1992) indicate that the \ha, \hb, and
\hg\ equivalent widths of a star are always the same to within 30 per
cent and do not depend sensitively on metallicity (see also 
Gonz\'alez-Delgado \& Leitherer 1999). We therefore adopt the \hg\
absorption equivalent widths listed in Table~2 of Cananzi et al. 
(1993) as estimates of the \ha\ and \hb\ equivalent widths of stars
of all metallicities. We compute the line equivalent widths in 
the integrated spectra of stellar populations by weighting the 
contributions from individual stars by their level of continuum 
(see, e.g., Bressan, Chiosi \& Tantalo 1996 for a description of
this standard procedure).

\subsection{Transmission by the Photoionized Gas}

We compute the transmission function $T_\lambda^{+}(t')$ of the gas
ionized by stars of age $t'$ (eqs.~[\ref{basic}]--[\ref{ttp}]) using 
the standard photoionization code CLOUDY (version C90.04; Ferland 1996). 
This requires us to specify the parameters of the gas. Here, we adopt
`effective' parameters to describe the ensemble of \hii\ regions 
and the diffuse gas ionized by all stars of age $t'$ within a galaxy.
This is motivated by the fact that the optical-line ratios in the 
integrated spectra of nearby spiral galaxies are similar to those
in the spectra of individual \hii\ regions (Kobulnicky et al. 1999;
see also \S3 below). Thus, photoionization models that reproduce
the properties of \hii\ regions should be adequate, modulo some 
adjustment of the parameters, to reproduce the integrated properties
of whole galaxies. For example, at fixed metallicity, the ratio of
ionizing-photon to gas densities is expected to be lower when averaged
over a whole galaxy than in individual \hii\ regions because of the 
contribution by diffuse ionized gas (e.g., Lehnert \& Heckman 1994; 
Martin 1997).

In the photoionization code CLOUDY, the gas is described as spherical
concentric layers centred on the ionizing source (assumed to be 
pointlike). The ratio between the radius of the innermost layer $r_{\rm
in}$ and the Str\"omgren radius $R_{\rm S}$ fixes the actual geometry
of a model and the ionization profile of the gas. The Str\"omgren radius
is defined by
\begin{equation}
R_{\rm S}^3 = 3 Q/(4\pi n_{\rm H}^2\epsilon\,\alpha_B)\,,
\label{rstrom}
\end{equation}
where $Q$ is the rate of ionizing photons, $n_{\rm H}$ the hydrogen density, 
$\epsilon$ the volume-filling factor of the gas (i.e., the ratio of
the volume-averaged hydrogen density to $n_{\rm H}$), and $\alpha_{\rm
B}$ the case-B hydrogen recombination coefficient (Osterbrock 1989). The
`ionization parameter' is defined as the ratio of ionizing-photon
to gas densities at the distance $r$ from the ionizing source,
\begin{equation}
U(r)=Q/(4\pi r^2 n_{\rm H}c)\,,
\label{ur}
\end{equation}
where $c$ is the speed of light. For $r_{\rm in}\ga R_{\rm S}$, the total 
thickness $\Delta r$ of the ionized region is much smaller than the 
Str\"omgren radius, $\Delta r\ll R_{\rm S}$, and the ionization parameter
is roughly constant throughout this region, $U(r)\approx U(r_{\rm in})$.
The geometry in this case is approximately plane-parallel, and the 
volume-averaged ionization parameter is $\langle{U}\rangle\approx 
U(r_{\rm in})$. For $r_{\rm in}\ll R_{\rm S}$, however, $U(r)$ is a strong
function of $r$, and the thickness of the ionized region approaches the
Str\"omgren radius, $\Delta r\sim R_{\rm S}$. The geometry in this case 
is truly spherical, and the volume-averaged ionization parameter is, assuming
constant $n_{\rm H}$,
\begin{equation}
\langle{U}\rangle\approx3Q/(4\pi R_{\rm S}^2 n_{\rm H} c) = 3 U(R_S)\,.
\label{umean}
\end{equation}
Substituting $R_{\rm S}$ from equation~(\ref{rstrom}) and neglecting the 
weak dependence of $\alpha_{\rm B}$ on $r$ through the electron temperature, 
this yields
\begin{equation}
\langle{U}\rangle\approx {{\alpha_B^{2/3}}\over{c}}
\left({{3Q \epsilon^2 n_{\rm H}}\over{4\pi}}\right)^{1/3}\,.
\label{uparam}
\end{equation}
We find that, at fixed \uva, our results are not sensitive to the assumed
geometry of the gas. For simplicity, we adopt $r_{\rm in}=0.01\,$pc and 
hence spherical geometry in all the following applications.

In our approach, the parameters of the photoionization code are effective 
ones, which describe the ensemble of \hii\ regions and the diffuse
gas ionized by a single stellar generation in a galaxy. Since the ionizing
radiation changes as the stars evolve, these effective parameters depend 
on time. We call \qav$(t')$ the effective rate of ionizing photons seen
by the gas irradiated by stars of age $t'$ throughout the galaxy. For
a fixed IMF, \qav$(t')$ defines a characteristic mass \mav\ through the 
relation
\begin{equation}
\tilde{Q}(t')= {{{\tilde{M}}_\ast}\over{hc}}\,
\int_0^{\lambda_L} d\lambda\,\lambda\,S_\lambda(t')\,,
\label{qmean}
\end{equation}
where $S_\lambda(t')$ is again the luminosity emitted per unit wavelength and
per unit mass by a stellar generation of age $t'$ ($\lambda_L= 912\,${\AA}).
The quantity \mav\ should be interpreted as the effective mass of the ionizing 
star clusters in the galaxy. For simplicity, we assume that the effective 
density \nav\ and the effective volume-filling factor \eav\ of the ionized gas
do not depend on $t'$. By analogy with equations~(\ref{ur})--(\ref{uparam}),
we define the effective ionization parameter of the gas irradiated by stars
of age $t'$ and its average over volume as
\begin{equation}
\tilde{U}(t',r)=\tilde{Q}(t')/(4\pi r^2 \tilde{n}_{\rm H}c)\,,
\label{urmean}
\end{equation}
\begin{equation}
\langle{\tilde{U}}\rangle(t')\approx3\;\tilde{U}(t',R_{\rm S})\approx 
{{\alpha_B^{2/3}}\over{c}}
\left[{{3\;\tilde{Q}(t')\;\tilde{\epsilon}^2\;\tilde{n}_{\rm
H}}\over{4\pi}}\right]^{1/3}\,.
\label{uparam2}
\end{equation}
Here $r$ and $R_{\rm S}$ pertain to the \hii\ region ionized by a star 
cluster of effective mass \mav. In our model, therefore, a galaxy 
containing several stellar generations comprises different gas components
having different effective ionization parameters. 

We must also specify the abundances of heavy elements and dust in
the gas. The `cosmic' abundances of heavy elements at solar 
metallicity are constrained to within only a factor of two by 
observations of Galactic stars (see the discussions by Garnett et 
al. 1995 and Snow \& Witt 1996). As a result, various assumptions 
can be found in previous photoionization studies about the cosmic 
abundances of important gas `coolants' such as N, O, and S (e.g., 
McCall, Rybski \& Shields 1985; Campbell 1988; McGaugh 1991; Shields
\& Kennicutt 1995; Stasi\'nska \& Leitherer 1996; Tresse et al. 1996;
Martin 1997; Bresolin, Kennicutt \& Garnett 1999).  Here, we adopt 
the following cosmic abundances for these elements by
number relative to hydrogen: 
(N/H)$_{\rm cosmic}=5.0\times10^{-5}$, 
(O/H)$_{\rm cosmic}=6.6\times10^{-4}$, 
and (S/H)$_{\rm cosmic}=1.3\times10^{-5}$.
These choices, which are within the ranges
considered in the above studies, were motivated in part by the desire to 
match roughly the observed typical (i.e., median) properties of the 
samples of nearby galaxies and \hii\ regions investigated in \S3. For 
all the other elements, we adopt the solar-composition abundances of 
Grevesse \& Anders (1989; as extended by Grevesse \& Noels 1993). 
Furthermore, for simplicity, we assume that the abundances of all 
elements except nitrogen scale linearly with the effective gas
metallicity, noted \zav. Observations of \hii\ regions in the Milky
Way and other nearby galaxies indicate that, at high metallicity, 
nitrogen is enhanced by secondary nucleosynthetic processing (e.g.,
Vila-Costas \& Edmunds 1993; Henry \& Worthey 1999, and references
therein). We assume that N/H scales as $\tilde{Z}^{2}$ for $\tilde{Z}
\ge0.4Z_\odot$, consistent with the observations compiled in Fig.~6
of Henry \& Worthey (1999). In all applications in this paper, we
adopt the same metallicity for the gas as for the stars (in practice,
we adopt the available stellar metallicity closest to that set
for the gas).

In our model, we treat the mass fraction of heavy elements locked
into dust grains in the ionized gas as a free parameter. We refer to 
this as the `effective dust-to-heavy element ratio', noted \xav. For
simplicity, we assume that, at fixed metallicity, the fraction of 
each refractory element depleted onto dust grains from the gas phase
scales linearly with \xav\ (the secondary production of non-refractory
nitrogen implies that, when \zav\ changes, this fraction scales almost
but not perfectly linearly with \xav). For all the elements except silicon,
we adopt the default depletion factors of CLOUDY for the average ISM
(Cowie \& Songaila 1986; Jenkins 1987). These factors correspond to
a default dust-to-heavy element ratio of 0.46 for the above cosmic
abundances. We adopt the more recent estimate of the silicon
depletion factor from {\it Hubble Space Telescope (HST)} observations
of \hii\ regions by Garnett et al. (1995). We take this to correspond
to a dust-to-heavy element ratio of 0.30, which is expected to be more
typical of \hii\ regions, where grains can be eroded and destroyed
(see Garnett et al. 1995; Shields \& Kennicutt 1995). The optical 
properties of the dust in the photoionization code are based on the 
standard Draine \& Lee (1984) grain model with extensions at ionizing
wavelengths by Martin \& Rouleau (1989). The resulting extinction curve
displays a maximum near 17~eV ($\lambda\approx730\,${\AA}). The most
important effect of dust in the photoionized gas is the depletion of 
coolants (Shields \& Kennicutt 1995). Other effects include the 
absorption and scattering by dust grains of the incident radiation and
the photoelectric heating and collisional cooling of the gas (see Ferland
1996 for more details).

Photoionization calculations are often parameterized in terms of the
ionization parameter. The reason for this is the similarity of the
results obtained for different combinations of $Q$, $\epsilon$, and 
$n_{\rm H}$ giving rise to the same $\langle U \rangle$ in 
equation~(\ref{uparam}) 
(e.g., Evans \& Dopita 1985; McCall et al. 1985; McGaugh 1991; Stasi\'nska
\& Leitherer 1996). By analogy, it is convenient here to fix \mav\ and 
hence \qav$(t')$ (eq.~[\ref{qmean}]) and use as a variable the value of
\uav$(t')$ at $t'=0$ (eq.~[\ref{uparam2}]). We refer to this quantity 
in the following as the `zero-age effective ionization parameter', noted
\begin{equation}
\langle{\tilde{U}}_0\rangle\equiv\langle{\tilde{U}}\rangle(0)\,.
\label{udef}
\end{equation}
Our results do not depend at all on the choice of \mav\ (\S3.1), except
through the fact that any \mav\ imposes a maximum \uavo\ at fixed \nav\ 
(eqs.~[\ref{qmean}] and [\ref{uparam2}]). For all but very high \uavo, 
therefore, we adopt \mav$ = 3\times10^4\,M_\odot$, corresponding to 
\qav$(0) \approx 9\times10^{50} \,$s$^{-1}$ for $m_{\rm U}=100\, M_\odot$
and $\tilde{Z} = Z_\odot$ (i.e., the equivalent of roughly 80 O7~V stars;
Vacca 1994). This is typical of the rate of ionizing photons in giant \hii\
regions in the Milky Way and other nearby galaxies (e.g., Kennicutt 1984;
for ${\rm \uavo} \ga10^{-1.5}$, we adopt \mav$=1\times10^6\,M_\odot$).
For given \nav\ and \uav, equations~(\ref{qmean}) and (\ref{uparam2}) 
allow us to compute \eav\ (we adopt $\alpha_B = 2.59 \times 10^{-13}
\, $cm$^3\, $sec$^{-1}$, appropriate for electronic temperatures T$_{\rm e}
\approx10^4\,$K; Osterbrock 1989). 

Throughout this paper, we assume that galaxies are ionization bounded. 
This is suggested by the upper limit of 3 per cent on the fraction of the
ionizing radiation escaping from nearby starburst galaxies (Leitherer et
al. 1995; see also Giallongo, Fontana \& Madau 1997). We stop the 
photoionization calculations when the electron density falls below 1 per
cent of the hydrogen density or if the temperature falls below 100~K. 
A model \hii\ region, as defined by equation~(\ref{urmean}), consists 
typically of an inner zone containing highly-ionized species (e.g., 
N$^{+2}$, O$^{+2}$), an intermediate zone (N$^{+}$, O$^{+}$), and an 
outer zone containing significant H$^0$ and other neutral and 
low-ionization species (O$^0$, S$^+$; see for example Fig.~1 of Evans 
\& Dopita 1985). The relative sizes of the different zones depend on the
model parameters.

\subsection{Emergent Luminosity}

We have now specified our model for computing the luminosity 
$L_\lambda^+(t)$ produced at the time $t$ by a star-forming galaxy
as defined by equation~(\ref{basic}). In summary, the main adjustable
parameters describing the luminosity $S_\lambda$ emitted by the stars
are: the upper cutoff mass of the IMF, $m_{\rm U}$; the star formation
rate, $\psi$; and the metallicity, $Z$. For the transmission function 
$T_\lambda^{+}$ of the photoionized gas (of effective metallicity 
$\tilde{Z}=Z$), the main adjustable parameters are: the zero-age 
effective ionization parameter, \uavo; the effective gas density, 
\nav;  and the effective dust-to-heavy element ratio, \xav. The other
main parameter is the age of the galaxy, $t$. In practice, we compute 
the transmission function $T_\lambda^{+}$ of the gas by subdividing
star formation into units of mass \mav. This amounts to dividing 
$\psi(t-t')$ and multiplying $S_\lambda(t')$ by \mav\ in 
equation~(\ref{basic}) and then using equation~(\ref{qmean}) to 
evaluate \qav$(t')$. 

Until now, we have ignored the absorption of photons emanating
from the ionized gas (and those emitted by old stars) by dust
in the `neutral ISM' before they escape from the galaxy. A 
detailed modelling of this effect is not necessary if the only 
observables under interest are the luminosities of optical 
emission lines, and if at least two H-recombination lines are
available. In this case, the observed line luminosities can be
corrected in a straightforward way for absorption by dust in the
neutral ISM. The reason for this is that the ionized gas always
produces H-recombination lines in ratios corresponding to
dust-free case~B recombination, whether or not dust is present
(Hummer \& Storey 1992; Ferland 1996; Bottorff et al. 1998; see
also \S4 below). The effect of dust is to reduce the ionization
rate and hence the luminosities of all H-recombination lines by a
similar amount. The departure of H-recombination line ratios
(e.g., \hahb) from dust-free case~B recombination is therefore an
efficient probe of dust in the neutral ISM of a galaxy. In addition,
since most known extinction curves (e.g., Milky Way, LMC, SMC, 
starburst galaxies) are similar at optical wavelengths, the 
absorption of other optical emission lines (e.g., \oii, \oiii,
\nii, \sii) can be inferred reliably from that of H-recombination
lines. In \S3.1, we use this remarkable property to calibrate
our model for $L_\lambda^+$.

A more refined description of the absorption by dust in the
neutral ISM is required to compute simultaneously the line and
continuum emission from galaxies. In nearby starburst galaxies,
the attenuation inferred from the \hahb\ ratio is typically
higher than that inferred from the spectral continuum (e.g.,
Fanelli, O'Connell \& Thuan 1988; Calzetti, Kinney \& 
Storchi-Bergmann 1994; Calzetti 1997). Recently, Charlot \& Fall
(2000) have developed a simple model for the absorption of starlight
by dust in galaxies, based on an idealized description of the main
features of the ISM. Charlot \& Fall show that the finite
lifetimes of the dense clouds in which stars form are a key 
ingredient for resolving the apparent discrepancy between the 
attenuation of line and continuum photons in nearby starburst
galaxies. Their model accounts for all the available observations
of these galaxies, including the ratio of far-infrared to
ultraviolet luminosities, the \hahb\ ratio, the \ha\ equivalent
width, and the ultraviolet spectral slope. We use this prescription
here to compute the absorption by dust in the neutral ISM.

The luminosity per unit wavelength $L_\lambda(t)$ emerging at
time $t$ from a star-forming galaxy can be inferred from the
expression of $L_\lambda^+(t)$ in equation~(\ref{basic}). We
write
\begin{equation}
L_\lambda(t)=\int_0^t\,dt'\,\psi(t-t')\,S_\lambda(t')
        \,T_\lambda^{+}(t')\,T_\lambda^{0}(t')\,,
\label{dustbasic}
\end{equation}
where $T_\lambda^{0} (t')$ is the transmission function of the 
neutral ISM. Charlot \& Fall (2000) show that the observed 
relationships between the various integrated spectral properties
of nearby starburst galaxies are well reproduced by the simple
recipe\footnote{The function denoted here by $T_\lambda^{0}(t')$
is equivalent to the function $T_\lambda(t')$ defined by equation~(6)
of Charlot \& Fall (2000) for $f=0.0$ in their notations.}
\begin{equation}
T_\lambda^{0}(t')=
\exp\left[-\hat{\tau}_\lambda^{\rm }(t')\right]\,,
\label{taudef}
\end{equation}
\begin{eqnarray}
\hat{\tau}_\lambda(t')=\cases{
\hskip0.20cm\hat{\tau}_V\left(\lambda/{5500\,{\rm \AA}}\right)^{-0.7}\,,
&for $t'\leq 10^7$ yr,\cr
{{\mu\hat{\tau}_V}}\left(\lambda/{5500\,{\rm \AA}}\right)^{-0.7}\,,
&for $t'>10^7$ yr\,.\cr}
\label{taueff}
\end{eqnarray}
The wavelength dependence of the `effective absorption' curve
$\hat\tau_\lambda$ is tightly constrained by an observed relation 
between ratio of far-infrared to ultraviolet luminosities and
ultraviolet spectral slope. The age $10^7\,$yr in the above expression, 
which coincides with the value of $t^{\rm OB}$ in our model (\S2.2),
corresponds to the typical timescale for young stars to disrupt their
dense `birth clouds' or migrate away from them into the `ambient ISM'.
Here, we have introduced for convenience the fraction $\mu$ of the total 
effective absorption optical depth in the neutral ISM (birth clouds 
plus ambient ISM) contributed by the ambient ISM. Charlot \& Fall 
(2000) show that $\mu \approx1/3$ approximates closely the observed 
relation between \hahb\ ratio and ultraviolet spectral slope for 
starburst galaxies. To account for the scatter about this relation,
we consider below values ranging from $\mu=1$ (i.e., all the radiation
from the \hii\ regions leaks into the ambient ISM) to $\mu=1/5$. We 
note that, for $\hat{\tau}_V=0$, equations~(\ref{basic}) and 
(\ref{dustbasic}) are equivalent, and $L_\lambda(t)=L_\lambda^{+}(t)$.

We compute the luminosity $L_\lambda(t)$ as the sum
\begin{equation}
L_\lambda(t)=L_\lambda^{\rm c}(t)+\sum_i \delta(\lambda-\lambda_i)L_i(t)\,,
\label{newbasic}
\end{equation}
where $L_\lambda^{\rm c}(t)$ is the luminosity per unit wavelength of 
the continuum radiation emerging from the galaxy (including the gas 
recombination continuum), $L_i(t)$ is the luminosity of the line 
emerging at the wavelength $\lambda_i$, and $\delta$ is the Dirac 
delta function. The equivalent width of an emergent line, noted 
$W_i(t)$, is the difference between the emission equivalent width 
[i.e., the ratio of $L_i(t)$ to the continuum luminosity near 
$\lambda_i$] and the stellar absorption equivalent width. In 
star-forming galaxies, stellar absorption is especially important
for H-recombination lines (see \S2.2). Finally, we relate the 
emergent luminosity of an emission line to the star formation rate
by the `line efficiency factor'
\begin{equation}
\eta_i(t) \equiv L_i(t)/\psi(t)\,.
\label{effpar}
\end{equation}
In the following, we are primarily interested in the dependence of
\etaha, \etaoii, and \etaoiii\ on the star-formation parameters of 
galaxies.

\section{Comparison with Observations}

In this section, we use observations of nearby star-forming galaxies to
calibrate our model. Our goal is to identify models that account for
the typical properties, scatter, and trends seen in the observations.
In \S3.1 below, we first describe the specific influence of each 
parameter in the model on the ratios of prominent optical emission 
lines. This allows us to identify how the \ha, \oii, and \oiii\ 
efficiency factors are related to observed line ratios. This analysis,
which involves line luminosities only, does not depend on our 
assumptions about the absorption by dust in the neutral ISM (\S2.4).
We must check, however, that the model also reproduces whole (line 
plus continuum) spectral energy distributions of nearby star-forming
galaxies. We show this in \S3.2, where we then have to specify the 
absorption by dust in the neutral ISM.

\subsection{Emission-Line Luminosities}

To calibrate our model, we compare it with observations of a sample
of nearby star-forming galaxies, for which the fluxes of the \ha, \hb,
\oii$\lambda{3727}$, \oiii$\lambda{5007}$, \nii$\lambda6583$, and 
\sii$\lambda\lambda{6717,\, 6731}$ emission lines are available. The
sample, which we have compiled from the literature, includes 92 non-Seyfert
galaxies spanning a wide range of morphological types, from Sab to Irr,
and a wide range of absolute $B$-band magnitudes, $-16\la M_B\la -22$
(we adopt a Hubble constant $H_0= 70 $~km$\,$s$^{-1}$Mpc$^{-1}$). 
About a third of these (32 galaxies) are normal spiral and irregular
galaxies from the sample of Kennicutt (1992b). Another third (33 
galaxies) are starburst galaxies selected by Calzetti, Kinney \& 
Storchi-Bergmann (1994) from the {\it International Ultraviolet 
Explorer} ({\it IUE}) atlas of Kinney et al. (1993), for which
optical-line fluxes are available from Storchi-Bergmann, Kinney \&
Challis (1995) and McQuade, Calzetti \& Kinney (1995). The remaining
third (27 galaxies) are low-metallicity, `\hii\ galaxies' (including
isolated extragalactic \hii\ regions and blue compact dwarf galaxies)
from the sample of Stasi\'nska \& Leitherer (1996). These include
17 galaxies with secure \ha\ measurements from the Terlevich et
al. (1991) catalogue and 10 galaxies observed by Izotov, Thuan \& 
Lipovetzky (1994). Most of the normal spiral and starburst galaxies
in our sample have \ha\ equivalent widths in the range 
10--150~{\AA}. The low-metallicity, \hii\ galaxies have typically 
much higher \ha\ equivalent widths, in the range 300--800~{\AA}.

We correct the flux ratios of all lines for absorption by dust in 
the neutral ISM on the basis of the observed \hahb\ ratios
(\S2.4). This requires us to first correct the observed \ha\
and \hb\ fluxes for stellar absorption. Calzetti et al. (1994)
and Izotov et al. (1994) derive \hb\ stellar absorption equivalent
widths for all the galaxies in their samples using the procedure
outlined by McCall et al. (1985), that is based on a simultaneous
fitting of the observed \ha, \hb, and \hg\ fluxes.  Also, 
Kennicutt (1992b) shows that a mean \hb\ absorption equivalent 
width of $-5$~{\AA} is appropriate for the galaxies in his sample, 
and Masegosa et al. (1994) suggest a mean \hb\ absorption equivalent
width of $-2$~{\AA} for the low-metallicity galaxies of the Terlevich
et al. (1991) catalogue (this is also roughly the mean value for
the galaxies of the Izotov et al. 1994 sample). To correct the 
observed \hahb\ ratio of each galaxy for stellar absorption, we
adopt the same absorption equivalent width for \ha\ as for \hb\ 
(see \S2.2 for a justification). We then evaluate the effective
absorption by dust in the neutral ISM by comparing the resulting \hahb\
ratio to the dust-free case~B recombination value of 2.86 (appropriate
for electronic densities $n_{\rm e}\la 10^4\,$cm$^{-3}$ and temperatures
$T_{\rm e}\approx 10^4\,$K; Osterbrock 1989). Finally, we deredden the 
relative ratios of the \oii, \hb, \oiii, \ha, \nii, and \sii\ lines 
using the Calzetti (1997) effective absorption curve (which is similar
to the Milky Way, LMC, and SMC extinction curves at optical wavelengths).
Neglecting possible anisotropies, we equate line flux ratios to the 
corresponding luminosity ratios. Figs.~1, 2, and 3 show \lpoiiihb\ as a
function of \lpoiioiii, \lpsiiha, and \lpniisii, respectively, for this 
sample. Evidently, the galaxies follow well-defined relations in
these diagrams. The typical measurement errors indicate that the scatter
about the mean relations is real.

\begin{figure}
%\vspace{302pt}
%\vspace{250pt}
\epsfxsize=9cm
\hfill{\epsfbox{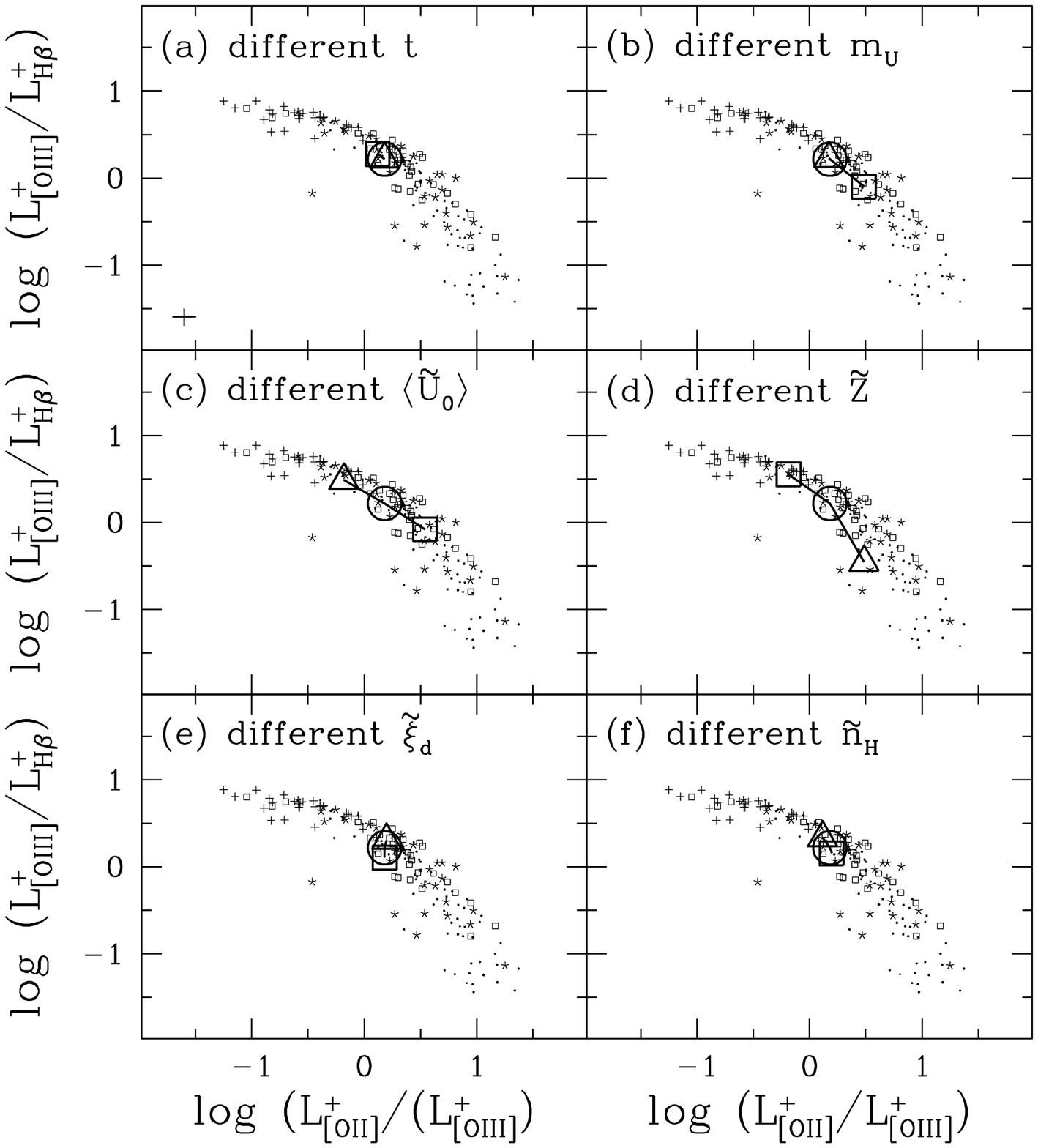}}
 \caption{\oiiihb\ ratio plotted against \oiioiii\ ratio. The 
absorption by dust in the neutral ISM and the stellar \hb\ absorption
are not included. The data points are from the samples of galaxies
and individual \hii\ regions discussed in \S3.1 and are repeated in all
panels ({\it squares}: normal spiral and irregular galaxies; {\it 
stars}: starburst galaxies; {\it crosses}: \hii\ galaxies; and {\it
dots}: \hii\ regions; typical measurement errors for galaxies are 
indicated in the upper left panel). In each case, the line shows the
effect of varying one parameter from the lower end of the range ({\it
square}) to the standard value ({\it circle}) to the upper end of the
range ({\it triangle}), with all other parameters fixed at their standard
values (eq.~[\ref{standard}]): ({\it a}) effective starburst age, $t=1
\times10^6$, $3\times10^8$, and $3\times10^9\,$yr; ({\it b}) upper
cutoff mass of the IMF, $m_{\rm U}= 40$, 100, and 130$\,M_\odot$;
({\it c}) zero-age effective ionization parameter: $\log({\rm \uavo})
= -3.0$, $-2.5$, and $-2.0$; ({\it d}) effective metallicity: ${\rm \zav}
=0.2Z_\odot$, $Z_\odot$, and $2Z_\odot$; ({\it e}) effective dust-to-heavy
element ratio: ${\rm \xav}= 0.1$, 0.3, and 0.5 ({\it f}) effective gas
density: ${\rm \nav}= 3$, 30, and 300~cm$^{-3}$.}
 \end{figure}

\begin{figure}
%\vspace{250pt}
\epsfxsize=9cm
\hfill{\epsfbox{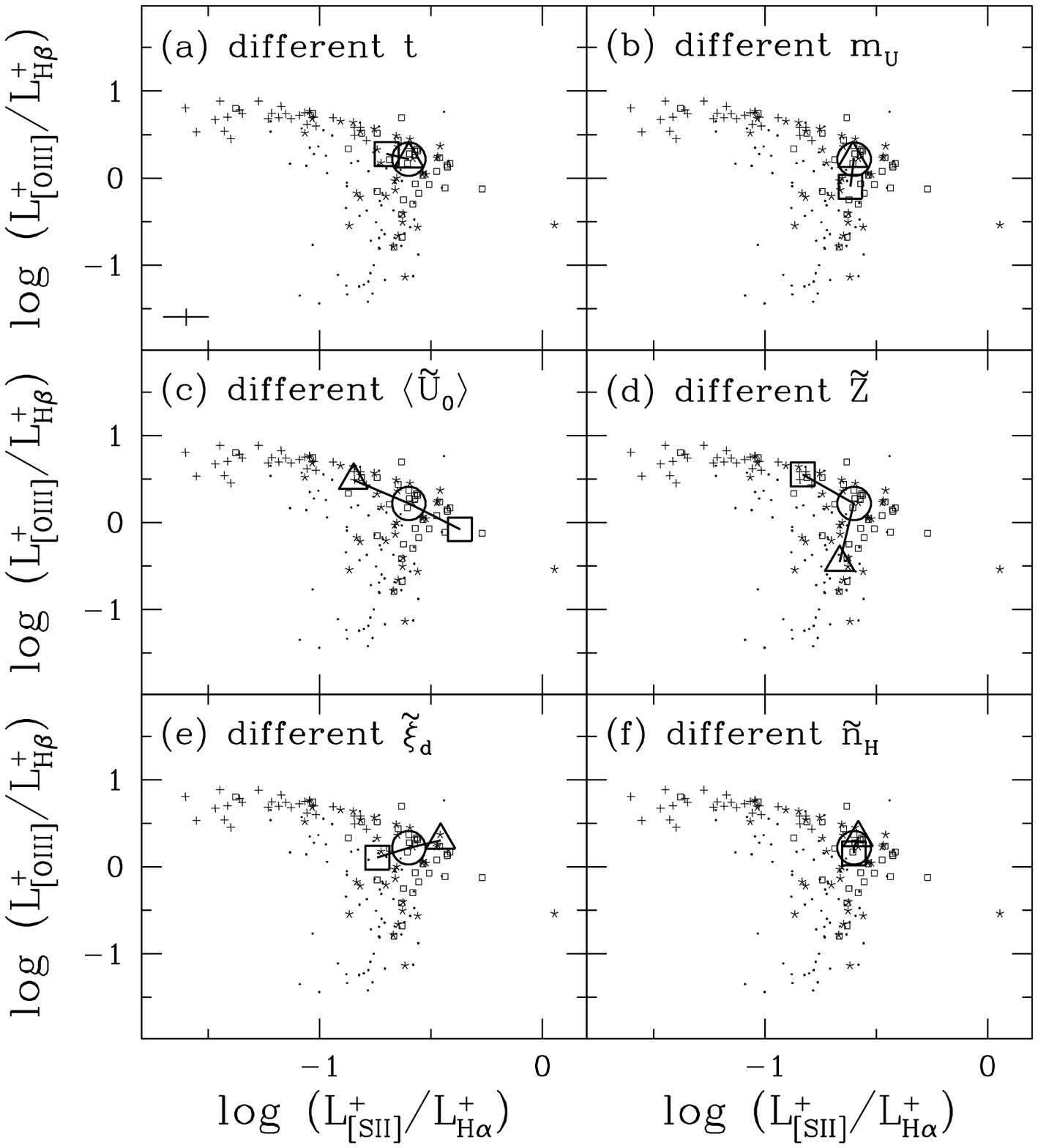}}
 \caption{\oiiihb\ ratio plotted against \siiha\ ratio. The absorption
by dust in the neutral ISM and the stellar \ha\ and \hb\ absorptions 
are not included. The data points are from the samples of galaxies and
individual \hii\ regions discussed in \S3.1 and are repeated in all
panels ({\it squares}: normal spiral and irregular galaxies; {\it
stars}: starburst galaxies; {\it crosses}: \hii\ galaxies; and {\it
dots}: \hii\ regions; typical measurement errors for galaxies are
indicated in the upper left panel). In each case, the line shows the
effect of varying one parameter from the lower end of the range ({\it
square}) to the standard value ({\it circle}) to the upper end of the
range ({\it triangle}), with all other parameters fixed at their 
standard values (eq.~[\ref{standard}]). The models are the same as in
Fig.~1.}
 \end{figure}

\begin{figure}
%\vspace{250pt}
\epsfxsize=9cm
\hfill{\epsfbox{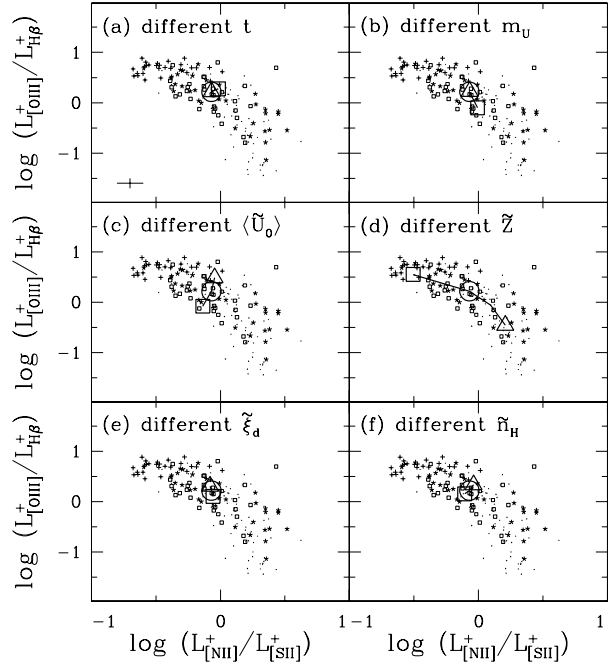}}
 \caption{\oiiihb\ ratio plotted against \niisii\ ratio. The absorption
by dust in the neutral ISM and the stellar \hb\ absorption are not
included. The data points are from the samples of galaxies and
individual \hii\ regions discussed in \S3.1 and are repeated in all
panels ({\it squares}: normal spiral and irregular galaxies; {\it
stars}: starburst galaxies; {\it crosses}: \hii\ galaxies; and {\it
dots}: \hii\ regions; typical measurement errors for galaxies are
indicated in the upper left panel). In each case, the line shows the
effect of varying one parameter from the lower end of the range ({\it
square}) to the standard value ({\it circle}) to the upper end of the
range ({\it triangle}), with all other parameters fixed at their 
standard values (eq.~[\ref{standard}]). The models are the same as in
Fig.~1.}
 \end{figure}

We now use these observations to constrain the parameters in our model.
Our goal is to identify models that account for the typical properties,
scatter, and trends seen in the sample in order to relate the \ha, \oii,
and \oiii\ efficiency factors to observed line ratios. The influence
of each parameter on observable quantities can best be explored by 
keeping all other parameters fixed at `standard' values.  For 
simplicity, we take for the moment the star formation rate $\psi$ to
be constant. In this case, the age $t$ of the stellar population should be 
regarded as the effective age of the most recent burst of star formation.
After some experimentation, we adopted the following standard parameters:
\begin{eqnarray}
t&=&3\times10^8\,{\rm yr}\cr
m_{\rm U}&=&100\,M_\odot\cr
\log({\rm \uavo})&=&-2.5\cr
{\rm \zav}&=&Z_\odot\cr
{\rm \xav}&=&0.3\cr
{\rm \nav}&=&30\,{\rm cm}^{-3}\,.
\label{standard}
\end{eqnarray}
While these values are not the result of a rigorous optimization procedure,
they do enable the standard model to match roughly the observed typical
(i.e., median) properties of the galaxy sample. We note that \xav$\,\approx
0.3$ is of the order of the dust-to-heavy element ratio expected in \hii\ 
regions (\S2.3). Also, ${\rm \nav}\approx30\,$cm$^{-3}$ is typical of the gas 
density in Galactic and extragalactic \hii\ regions (10--100~cm$^{-3}$;
Osterbrock 1989). For this \nav, $\log({\rm \uavo})\approx-2.5$ implies ${\rm
\eav}\approx 0.04$ (\S2.4), which is typical of the volume-filling factor in
giant \hii\ regions in the Milky Way and other nearby galaxies (e.g., 
Kennicutt 1984).

\begin{figure}
%\vspace{250pt}
\epsfxsize=9cm
\hfill{\epsfbox{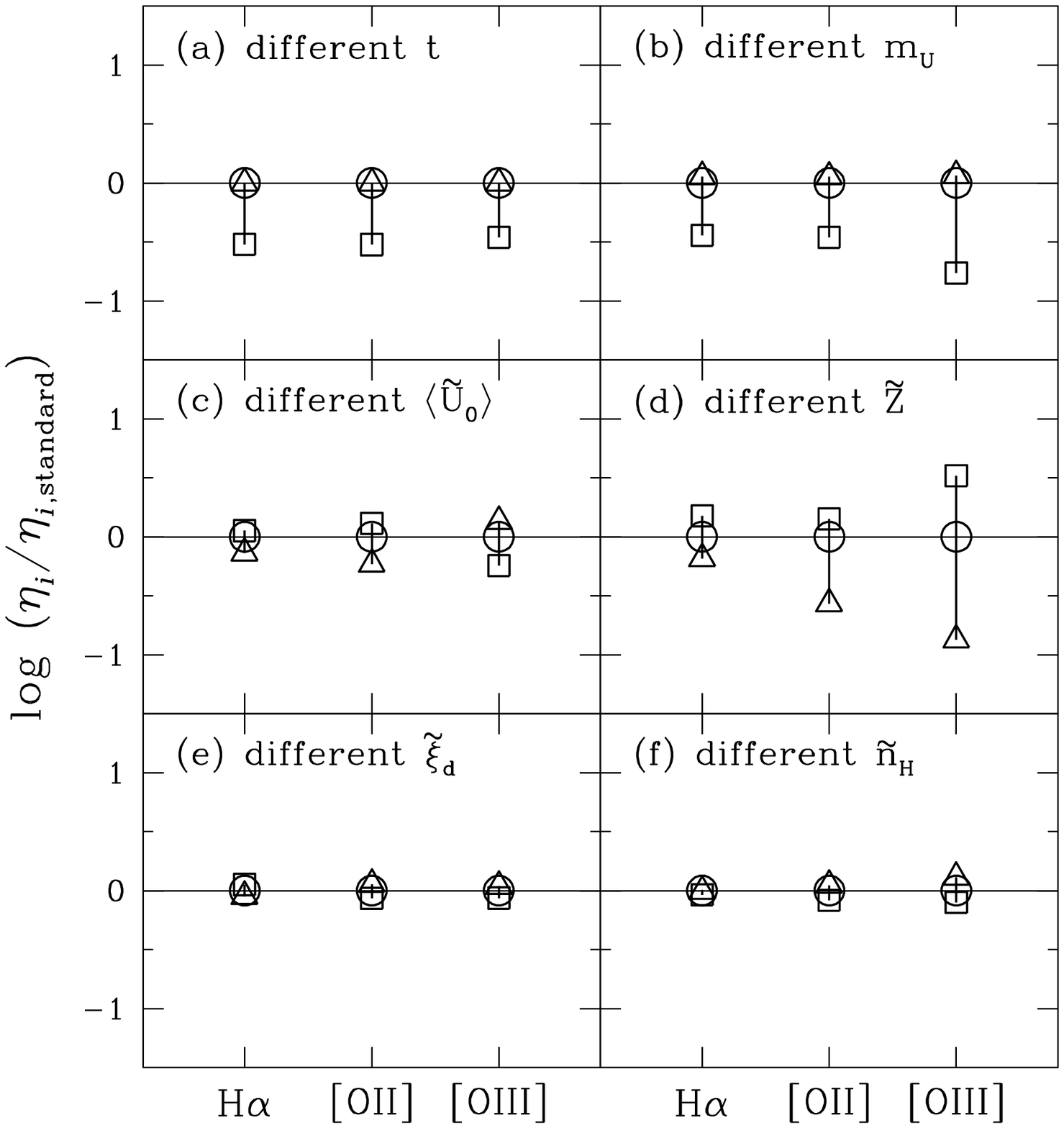}}
 \caption{\ha, \oii, and \oiii\ efficiency factors (as defined by 
eq.~[\ref{effpar}]) normalized to the standard model of eq.~[\ref{standard}].
The absorption by dust in the neutral ISM and the stellar \ha\ absorption
are not included. In each panel, the line shows the effect of varying
one parameter from the lower end of the range ({\it square}) to the 
standard value ({\it circle}) to the upper end of the range ({\it 
triangle}), with all other parameters fixed at their standard values
(eq.~[\ref{standard}]). The models are the same as in Fig.~1.}
 \end{figure}

We now compare the predictions of our model with the observations described
above. Each panel in Figs.~1--3 shows the effect of increasing and decreasing
one parameter with respect to its standard value with the others held fixed.
In Fig.~4, we show the influence of these variations on the \ha, \oii, and 
\oiii\ efficiency factors \etaha, \etaoii, and \etaoiii\ computed using 
equation~(\ref{effpar}) with $\hat\tau_V=0$ (i.e., for $L_i=L_i^+$). We can
summarize the role of each parameter as follows.
\begin{description}
\item {\it Effective starburst age}$\,$. Since we have assumed $\psi={\rm 
const}$, increasing $t$ makes \etaha, \etaoii, and \etaoiii\ larger initially,
as massive stars with lifetimes 3--5$\times 10^6\,$yr accumulate on the
main sequence (Fig.~4{\it a}). Stars with ages between 5$\times 10^6\,$yr 
and $t^{\rm OB} =1 \times10^7 \,$yr produce less ionizing photons, while
older stars do not contribute to the ionizing radiation in our model (\S2).
After $1\times10^7 \,$yr, therefore, a steady population of \hii\ regions with
effective ionization parameters ranging from \uavo\ to $\langle{\tilde{U}}\rangle
(t^{\rm OB})$ is established (eq.~[\ref{uparam2}]). The rising contribution
of gas with $\langle{\tilde{U}} \rangle(t)<\langle{ \tilde{U_0}} \rangle$
at early ages has a small but significant effect on the line ratios in 
Figs.~1{\it a}--3{\it a}, which can be understood from our discussion of
\uavo\ below.  

\item {\it Upper cutoff mass of the IMF}$\,$. Increasing $m_{\rm U}$ makes
\etaha, \etaoii, and \etaoiii\ larger because, near the cutoff, more massive
stars produce more ionizing photons. This effect becomes negligible for 
$m_{\rm U}>100\, M_\odot$ (Fig.~4{\it b}). Increasing $m_{\rm U}$ and hence
\qav$(t')$ (eq.~[\ref{qmean}]) also makes the effective ionization parameter
larger in the inner (O$^{+2}$) parts of the \hii\ regions (eq.~[\ref{urmean}]). 
At fixed \uavo, however, the effective ionization parameter in the outer, 
low-ionization (S$^+$) boundaries near the Str\"omgren radius is also 
fixed (eq.~[\ref{uparam2}]). Thus, increasing $m_{\rm U}$ makes the \oiiihb\
ratio larger and the \oiioiii\ ratio smaller but affects weakly the \siiha\ 
and \niisii\ ratios (Figs.~1{\it b}--3{\it b}; see \S2.3).

\item {\it Zero-age effective ionization parameter}$\,$. Increasing \uavo,
which corresponds here to increasing \eav\ at fixed \nav\ and \qav$(0)$,
causes more gas to be concentrated in the inner, high-ionization 
parts of the \hii\ regions; thus the \oiiihb\ ratio and \etaoiii\ increase,
and the \oiioiii\ ratio and \etaoii\ decrease (Figs.~1{\it c} and 4{\it c}).
Increasing \uavo\ also increases the effective ionization parameter near
the Str\"omgren radius ($R_{\rm S}$) and weakens the low-ionization \sii\ 
line (Figs.~2{\it c} and 3{\it c}). Since the H-column density scales 
roughly as $\tilde{\epsilon} \,\tilde{ n}_{\rm H}\,R_{\rm S}\propto 
\langle{ \tilde{U}} \rangle$, increasing \uavo\ at fixed dust-to-gas 
ratio \xav\zav\ causes more ionizing photons to be absorbed by dust and
hence reduces \etaha\ (Fig.~4{\it c}).

\item {\it Effective Metallicity}$\,$. Increasing \zav\ makes the 
cooling by heavy elements more efficient and reduces the electronic
temperature.  The cooling becomes dominated by infrared fine-structure
transitions with low excitation energies, and fewer electrons are capable
of exciting optical lines (e.g., Spitzer 1978). Thus, the \oiiihb\ 
ratio and \etaoii\ and \etaoiii\ decline (Figs.~1{\it d} and 4{\it d}).
The \oiioiii\ ratio increases in Fig.~1{\it d} because of the efficient
fine-structure cooling by doubly-ionized species in the inner parts
of the \hii\ regions (Stasi\'nska 1980). The \siiha\ ratio, which
increases initially due to the higher S$^{+}$ abundance, also 
declines when cooling becomes important (Fig.~2{\it d}). In Fig.~3{\it
d}, the increase of the \niisii\ ratio with \zav\ follows from our 
inclusion of nitrogen from secondary processing (\S2.3). We note that
the rate of ionizing photons produced by massive stars tends to decrease
with increasing metallicity (e.g., Fig.~4 of Garc\'{\i}a-Vargas et al.
1995). This effect accounts for roughly two thirds of the drop in \etaha\
as \zav\ increases in Fig.~4{\it d}, one third arising from the
larger fraction of ionizing photons absorbed by dust as \xav\zav\ 
increases.

\item {\it Effective dust-to-heavy element ratio}$\,$. Increasing \xav\
depletes important coolants such as oxygen from the gas phase. As the 
cooling through the infrared fine-structure transitions of these elements 
declines, the electronic temperature increases and makes cooling through
optical transitions more efficient (Shields \& Kennicutt 1995). The 
\oiiihb\ ratio and \etaoii\ and \etaoiii\ increase only slightly in 
Figs.~1{\it e} and 4{\it e} because the rise in $T_{\rm e}$ is compensated
by the depletion of oxygen from the gas phase. However, since S and N are
not refractory elements, the \siiha\ ratio increases substantially with
\xav\ at roughly constant \niisii\ ratio (Figs.~2{\it e} and 3{\it e}).
Increasing \xav\ also causes more ionizing photons to be absorbed by dust
and hence reduces \etaha\ (Fig.~4{\it e}).

\item {\it Effective gas density}$\,$. Increasing \nav\ makes collisional
de-excitation compete with radiative cooling. Since the infrared 
fine-structure transitions have lower critical densities for collisional
de-excitation than the optical transitions, their cooling efficiency 
declines when increasing \nav\ (Stasi\'nska 1990). The effect is 
negligible at low metallicities but more significant at high 
metallicities,  where the fine-structure lines are the dominant 
coolants (Oey \& Kennicutt 1993). The resulting rise in $T_e$ makes
the cooling through optical transitions more important; hence \etaoii\ 
and \etaoiii\ increase slightly (Figs.~4{\it f}). Changes in \nav\ have
a negligible effect on \etaha\ and a small effect on the line ratios in
Figs.~1{\it f}--3{\it f}.

\item {\it Other parameters}$\,$. We have also computed models with 
different IMF slopes (not shown). Changes in $x$ have a negligible
effect on the line ratios, which are determined by the properties of 
the most massive stars. For a power-law IMF, however, increasing $x$
causes more mass to be locked in low-mass stars and reduces the line
efficiency factors. We also tested the influence of changing the
effective mass of the ionizing star clusters and found that models
with different \mav\ and fixed \uavo\ are virtually identical.
\end{description}

Since each parameter in our model has a specific influence on the
ratios of prominent optical emission lines in star-forming galaxies,
we can comment on the physical origin of the relations defined by
these observed quantities. Figs.~1--3 show that the observed mean
trends can be most naturally reproduced by a sequence in the 
effective gas metallicity, although a spread in the effective 
ionization parameter is also indicated. In fact, at one end of the
observed relations, the very low \oiioiii\ ratios of the \hii\ 
galaxies with the smallest \niisii\ ratios appear to require high
effective ionization parameters in addition to low metallicities 
(Izotov et al. 1994 estimate ${\rm \zav}\approx 0.1 Z_\odot$ from a
detailed spectral analysis of these galaxies; this is close to the
value ${\rm \zav}=0.2 Z_\odot$ of the lowest-metallicity model in
Figs~1{\it d}--3{\it d}).  At the other end, the \oiiihb, \oiioiii,
and \niisii\ ratios of the most extreme normal spiral and starburst
galaxies appear to require low effective ionization parameters in
addition to high metallicities. As Figs.~1--3 indicate, the 
intrinsic scatter about the observed relations can arise from 
variations in the effective dust-to-heavy element ratio, the effective
gas density, and the effective age of the starburst (and possibly the
IMF).

The primary dependence of optical emission-line ratios on the 
metallicity and ionization parameter of the gas is a well-known
property of individual \hii\ regions (e.g., Baldwin, Phillips \&
Terlevich 1981; Evans \& Dopita 1985; McCall et al. 1985; Campbell
1988; McGaugh 1991; Shields \& Kennicutt 1995; Bresolin et al.
1999). To illustrate the parallel between galaxies and individual
\hii\ regions, we also show in Figs.~1--3 observations of 68 \hii\
regions in nearby spiral and irregular galaxies from McCall et al.
(1985). The observed relations for \hii\ regions are similar to 
those for galaxies.  This similarity, which justifies a posteriori 
our adoption of effective parameters to describe the nebular properties
of whole galaxies, arises presumably from the limited range of 
metallicities and ionization parameters of the gas in most 
star-forming galaxies (e.g., Zaritsky, Kennicutt \& Huchra 1994;
Kobulnicky et al. 1999). The rough coincidence between starburst
and normal spiral galaxies in Figs.~1--3 further suggests that 
stochastic bursts of star formation do not strongly influence the
global gas parameters of galaxies.

Fig.~2 does reveal a difference between the nebular properties of 
galaxies and those of individual \hii\ regions. At fixed \oiiihb\ 
ratio, galaxies tend to have higher \siiha\ ratios than individual 
\hii\ regions. This enhancement in the relative luminosity of the
low-ionization \sii\ line is thought to be a signature of the diffuse
ionized gas in galaxies (e.g., Lehnert \& Heckman 1994; Martin 1997;
Wang et al. 1997). It is usually associated with enhanced \oi$\lambda
6300$, \nii, and \oii\ emission and reduced \oiii\ emission. 
Fig.~2{\it c} confirms that the enhanced \siiha\ ratios of galaxies 
relative to individual \hii\ regions can be interpreted in terms
of a lower effective ionization parameter at fixed effective 
metallicity. We note, however, that the effective dust-to-heavy element
ratio in the ionized gas has also a major influence on the \siiha\ 
ratio (Fig.~2{\it e}). While decreasing \uavo\ makes the \oiiihb\ 
ratio smaller, increasing \xav\ allows models to reproduce the
observations of galaxies with both large \siiha\ and \oiiihb\
ratios (the effects of \xav\ on the \niiha\ and \oiha\ ratios are
similar to those on the \siiha\ and \oiiihb\ ratios, respectively).
It is worth recalling that, in our model, the details of the \hii\
regions and the diffuse gas ionized by each stellar generation in a
galaxy are subsumed in our description of the gas by effective 
parameters.

One of the most important novelties of our model is that it 
allows us to relate line efficiency factors to observed 
optical-line ratios. This has fundamental implications for SFR
estimates, as can be illustrated by the following example. Figs.~1
and 4 show that increasing \uavo\ or decreasing \zav\ relative to our 
standard model have similar effects on the observed \oiioiii\ and
\oiiihb\ ratios, but opposite effects on \etaoii. Thus, for fixed
\oiioiii\ and \oiiihb\ ratios, there can be a spread of at least a
factor of several in \etaoii, depending on the effective metallicity
and effective ionization parameter of the gas. Fig.~3 shows, however,
that the \niisii\ ratio can help us distinguish between variations
in \uavo\ and \zav. Therefore, by combining information about the
\oiiihb, \oiioiii, and \niisii\ ratios, we can reduce the uncertainties
in \etaoii\ and hence better constrain the star formation rate. We
exploit this property of our model in \S4 below, where we construct 
estimators of the star formation rate for different assumptions about
the available spectral information.

\subsection{Spectral Energy Distributions}

Up to now, we have compared our model with observations only
in terms of emission-line luminosities. Since we compute in a
consistent way the emission from stars and gas, it is important
to check that the model can also reproduce whole (line plus 
continuum) spectral energy distributions of observed 
galaxies. This requires us to consider two complications: the
contamination of H-recombination lines by stellar absorption;
and the absorption of line and continuum photons by dust in 
the neutral ISM. In \S2, we described how these effects are 
accounted for in our model.

We now compare our model with observed spectral energy 
distributions of various types of nearby star-forming galaxies. 
The integrated spectra at wavelengths 3650--7100~{\AA} of the 
galaxies in the Kennicutt (1992b) sample used in \S3.1 are 
available from Kennicutt (1992a). To avoid using the spectrum
of any arbitrary galaxy as reference, we extract observed 
`templates' from this sample by averaging the spectra of 
a few galaxies of the same morphological type with roughly
similar emission-line properties. Figs.~5, 6, and 7 show 
such spectral templates for Sb, Sc, and (Magellanic-irregular)
Sm/Im galaxies, respectively. Each spectrum is an average over
three galaxies with comparable \oiioiii\ and \hahb\ ratios.
For reference, the \oiioiii\ ratios and the \ha\ and \hb\ 
equivalent widths of the Sb, Sc, and Sm/Im templates are $L_{\rm
[O\,{\sc ii}]}/ L_{\rm [O\, {\sc iii}]}\approx 3.3$, 2.6, and
1.4, $W_{{\rm H} \alpha} \approx 14$, 26, and 82~{\AA}, and 
$W_{{\rm H}\beta} \approx-0.7$, 3.0, and 14~{\AA}, respectively.
The spectral resolution is about 800.

To reproduce these observed templates, we use various spectral
features to adjust the different parameters in our model. The
equivalent widths of emission lines and the stellar continuum
(in particular, the 4000~{\AA} discontinuity) constrain the
ratio of young (blue) to old (red) stars in the model and hence
the star formation history (e.g., Kennicutt 1983; Bruzual \&
Charlot 1993). In principle, these constraints would depend on
the unknown metallicity and the unknown dust content of the 
galaxies. However, since the star formation history fixes the 
stellar \ha\ and \hb\ absorptions, the \hahb\ ratio provides an
independent measure of the absorption by dust. Then, using the
\niisii\ and \oiioiii\ ratios, we can adjust the effective 
metallicity and the effective ionization parameter of the gas
(Figs.~1{\it c} and 3{\it d}). Finally, for fixed values of 
these parameters, the \siiha\ ratio constrains the dust content
of the ionized gas (Fig.~2{\it e}). Therefore, we can tune in
a methodic way the parameters of our model to reproduce the 
observations.

\begin{figure}
%\vspace{250pt}
\epsfxsize=9cm
\hfill{\epsfbox{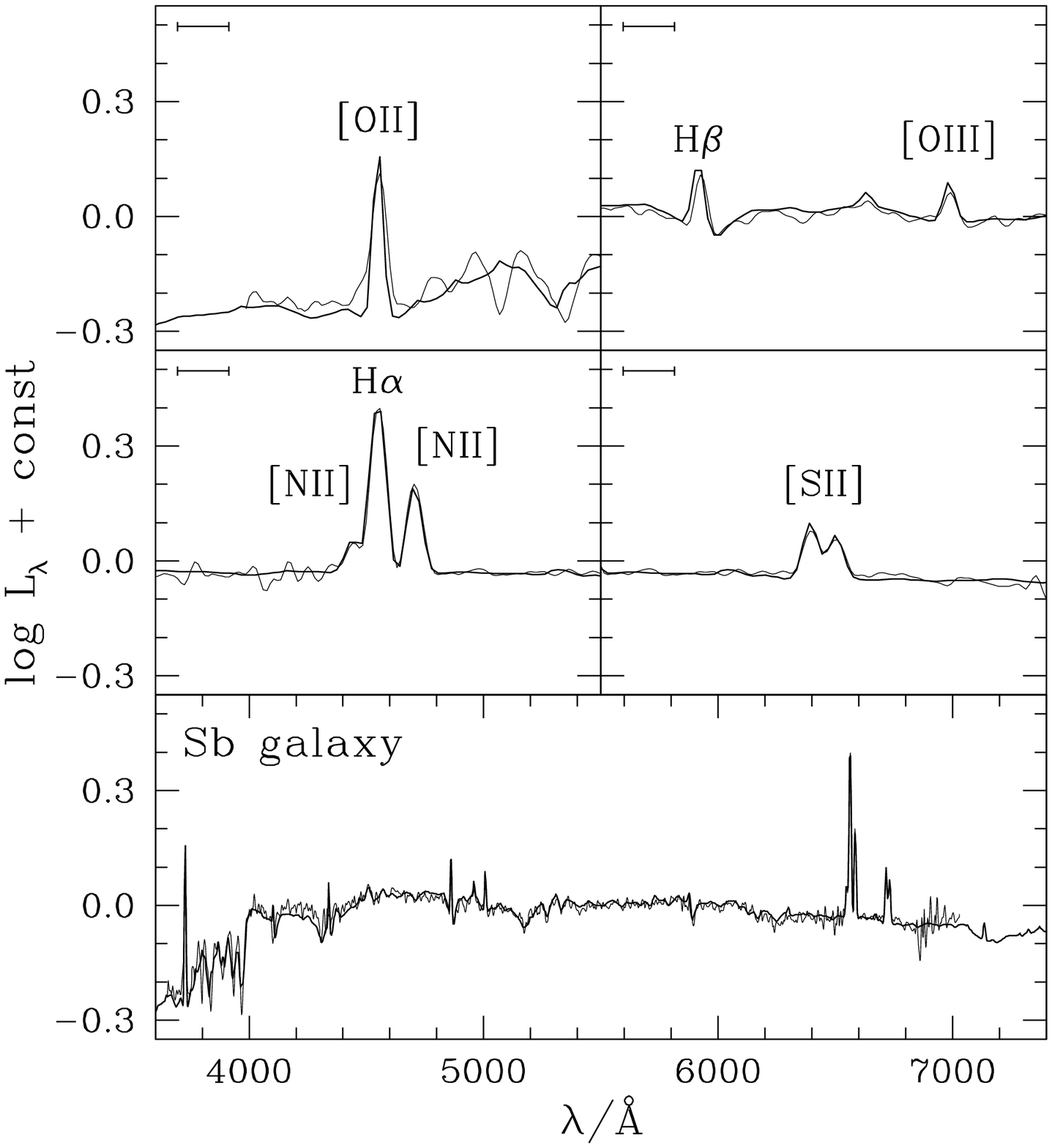}}
 \caption{Spectral fit of an observed Sb galaxy template. 
The template ({\it thin line}) is an average over three Sb 
galaxies with comparable \oiioiii\ and \hahb\ ratios in
the Kennicutt (1992a) atlas (NGC~1832, NGC~3147, and NGC~3627).
The model ({\it thick line}) has a exponentially declining 
star formation rate (timescale 5~Gyr) and an age $t\approx
9$~Gyr. The gas parameters are: $\log({\rm \uavo})= -3.0$, 
${\rm \zav}=1.5Z_\odot$, ${\rm \xav}=0.3$, and $\hat{\tau}_V
=1.0$. The upper four panels show the most prominent lines on
a larger wavelength scale (a mark indicates the scale 
$\Delta\lambda= 30$~{\AA}).}
 \end{figure}

\begin{figure}
%\vspace{250pt}
\epsfxsize=9cm
\hfill{\epsfbox{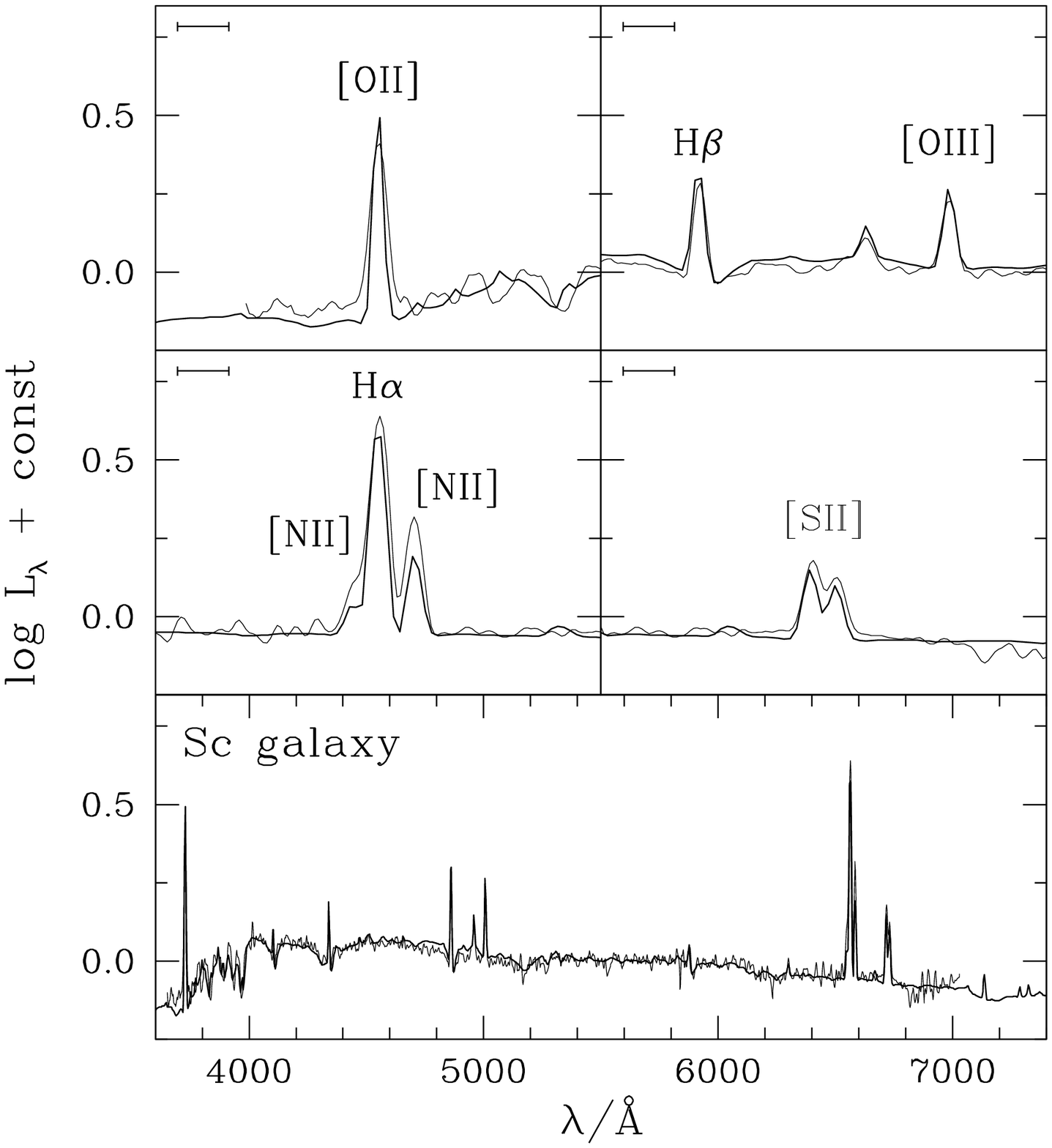}}
 \caption{Spectral fit of an observed Sc galaxy template. 
The template ({\it thin line}) is an average over three Sc 
galaxies with comparable \oiioiii\ and \hahb\ ratios in
the Kennicutt (1992a) atlas (NGC~2276, NGC~2903, and NGC~6181).
The model ({\it thick line}) has a exponentially declining 
star formation rate (timescale 15~Gyr) and an age $t\approx
8$~Gyr. The gas parameters are: $\log({\rm \uavo})= -3.0$,
${\rm \zav}=Z_\odot$, ${\rm \xav}=0.1$, and $\hat{\tau}_V=0.8$.
The upper four panels show the most prominent lines on a larger 
wavelength scale (a mark indicates the scale $\Delta\lambda=
30$~{\AA}).}
 \end{figure}

\begin{figure}
%\vspace{250pt}
\epsfxsize=9cm
\hfill{\epsfbox{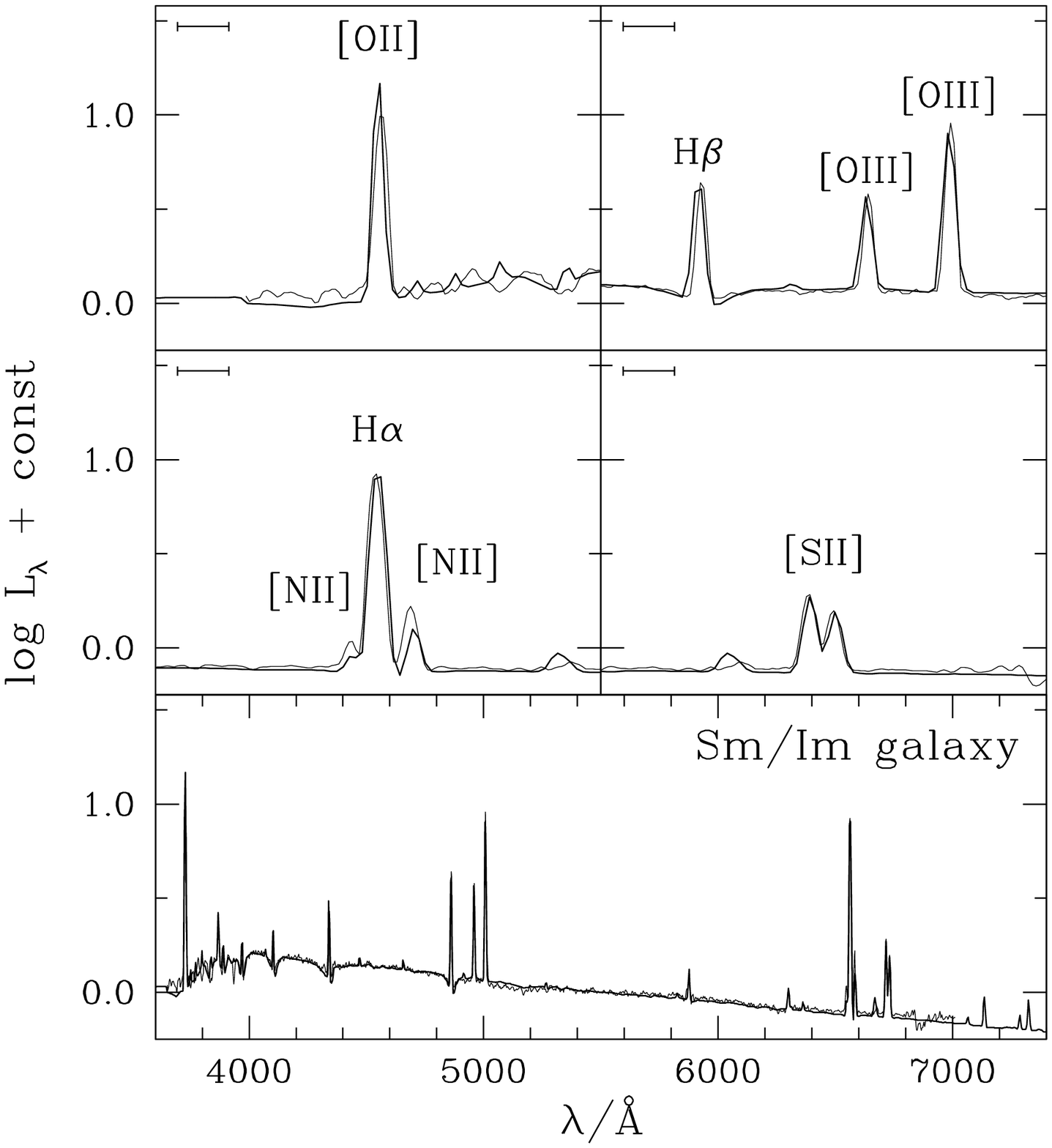}}
 \caption{Spectral fit of an observed Sm/Im galaxy template. 
The template ({\it thin line}) is an average over three 
(Magellanic-irregular) Sm/Im galaxies with comparable
\oiioiii\ and \hahb\ ratios in the Kennicutt (1992a) atlas
(NGC~4449, NGC~4485, and NGC~4670). The model ({\it thick line})
has a constant star formation rate and an age $t\approx2.5$~Gyr.
The gas parameters are: $\log({\rm \uavo})= -3.0$, ${\rm \zav}=
0.2Z_\odot$, ${\rm \xav}=0.1$, and $\hat{\tau}_V=0.1$. The upper
four panels show the most prominent lines on a larger wavelength
scale (a mark indicates the scale $\Delta\lambda=30$~{\AA}).}
 \end{figure}

\begin{figure}
%\vspace{250pt}
\epsfxsize=9cm
\hfill{\epsfbox{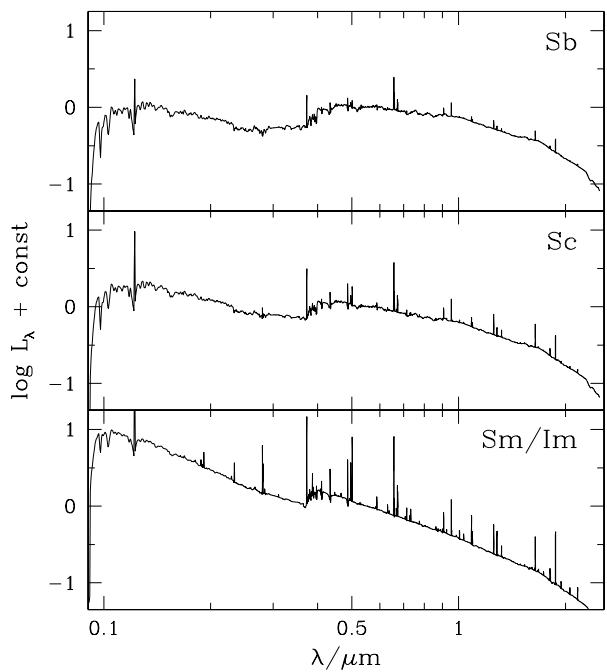}}
 \caption{Spectral energy distributions of the same models
as in Figs.~5--7 shown with extensions at ultraviolet and 
near-infrared wavelengths.}
 \end{figure}

To select models that are consistent with the observations,
we explore only a coarse grid of parameters (see Table~\ref{partable}
below). Attempting to fine tune the parameters might be of
limited validity because the observed spectral templates 
were built from individual galaxies that may have slightly 
different properties. We take the star formation rate $\psi(t)$
to be either constant or exponentially declining. In practice,
the results are mostly sensitive to the ratio of the current 
SFR to the average past SFR. We also explore variations in
the zero-age effective ionization parameter, \uavo, the effective
metallicity, \zav, the effective dust-to-heavy element ratio in 
the ionized gas, \xav, and the effective absorption optical depth
at 5500~{\AA} in the neutral ISM, $\hat{\tau }_V$ 
(eqs.~[\ref{taudef}] and [\ref{taueff}]). We fix the other 
parameters at their standard values, i.e.,  $\mu=1/3$ 
(eq.~[\ref{taueff}]), $m_{\rm U}=100 \,M_\odot$, and $\tilde{n
}_{\rm H}=30\,{\rm cm}^{-3}$ (eq.~[\ref{standard}]). Figs.~5,
6, and 7 show models selected in this way that reproduce the 
observed Sb, Sc, and Sm/Im spectral templates. In each case, the
continuum spectral energy distribution and the luminosities of
all the optical emission lines in the models agree well with the
observations. Given the idealizations of the models, we consider
these results as very satisfactory. The line strengths in the 
models are also quantitatively close to those quoted above for 
the observed spectral templates. For reference, the model Sb, 
Sc, and Sm/Im spectra, respectively, have $L_{\rm [O\,{\sc 
ii}]}/L_{\rm [O\, {\sc iii}]} \approx 2.9$, 2.5, and 1.6, 
$W_{{\rm H} \alpha} \approx 14$, 30, and 89~{\AA}, and 
$W_{{\rm H}\beta} \approx-0.4$, 2.3, and 14~{\AA}. The 
contribution to these equivalent widths by stellar absorption
is $-5.2$, $-6.2$, and $-7.3$~{\AA}, respectively.

These spectral fits constrain simultaneously the stars, gas, 
and dust parameters of the galaxies. We find that the ratio
of the current SFR to the average past SFR favored by
the observations ranges from 0.4 for the Sb galaxy template
to 0.8 for the Sc galaxy template to 1.0 for the Sm/Im
galaxy template. This is consistent with the expected dependence
of this ratio on morphological type (e.g., Kennicutt, Tamblyn
\& Congdon 1994). Furthermore, the emission-line ratios
of the Sb and Sc galaxy templates are well reproduced by models 
with roughly solar effective metallicity and $\hat{\tau 
}_V\approx1$, while the (Magellanic-irregular) Sm/Im galaxy
template is better fitted by a model with both low effective
metallicity and dust content. All three spectra favor the same
zero-age effective ionization parameter, ${\rm \uavo}\approx
-3.0$, but slightly different effective dust-to-heavy element
ratios in the ionized gas. Of course, these are only rough 
constraints on the star formation properties of the observed 
galaxies, as we have deliberately avoided fine tuning the model
parameters (the uncertainties in the derivation of the physical
parameters of galaxies will be addressed in \S4). In addition, the 
observed templates in Fig.~5--7 are representative of
only some Sb, Sc, and Sm/Im galaxies (see above). Thus, these
results should be taken merely as illustrative of the ability
with our model to constrain simultaneously the stars, gas, and
dust parameters of galaxies using integrated spectral energy
distributions.

We have described here the optical properties of our model for
the emission from star-forming galaxies. In reality, we compute
spectral energy distributions of the line and continuum emission
over the entire range of wavelengths from 91~{\AA} to 5~$\mu$m
(redward of this limit, the emission from polycyclic aromatic 
hydrocarbons, which are not included in the model, becomes 
important). As an example, Fig.~8 shows extensions of the model
spectra in Figs.~5--7 at ultraviolet and near-infrared wavelengths.
The prominent emission lines outside the optical window (e.g.,
\mgii\ at $\lambda=2798\,${\AA} and \siii, \feii, P$\alpha$ at
near-infrared wavelengths) offer complementary diagnostics of the 
star formation parameters, which we plan to study in future work.
Finally, it is worth recalling that the prescription we have adopted
to describe the absorption by dust in the neutral ISM ensures 
that our model is also consistent with the observed ultraviolet and 
far-infrared properties of nearby starburst galaxies (see \S2.4 above). 

\section{Observational Constraints on the Stars, Gas, and Dust
Parameters of Galaxies}

We now have a model for relating the integrated spectral properties
of galaxies to physical parameters such as the star formation rate,
the metallicity, and the effective absorption optical depth of the
dust. In practice, whole spectral energy distributions of the type
analyzed in \S3.2 are not always available to constrain these 
parameters in galaxies. However, the tight correlations between 
the different integrated spectral properties of nearby star-forming
galaxies suggest that even partial spectral information may be 
useful to constrain physical parameters (Figs.~1--3 above; see
also Figs.~2 and 3 of Charlot \& Fall 2000).  Here we analyze these 
observed relations with our model to construct estimators of the
star formation rate, the gas-phase oxygen abundance, and the 
effective absorption optical depth of the dust in galaxies for 
various assumptions about the available spectral information.

Our first step is to select a comprehensive set of models that
account for the full range of observations of nearby star-forming 
galaxies. We use the observed emission-line ratios of the galaxies in
Figs.~1--3 to identify the appropriate ranges in zero-age effective
ionization parameter, \uavo, effective metallicity, \zav, and effective 
dust-to-heavy element ratio, \xav. Since the effective gas density has
only a weak influence on line luminosities (Figs.~1{\it f}--4{\it f}),
we fix this parameter at its standard value of equation (\ref{standard}),
i.e., $\tilde{n }_{\rm H}=30\,$cm$^{-3}$. We also consider variations
in the total effective absorption optical depth of the dust in the 
neutral ISM, $\hat{\tau}_V$, and the fraction of this contributed by
the ambient ISM, $\mu$ (eq.~[\ref{taueff}]). These are constrained
by the observed relations between ratio of far-infrared to ultraviolet
luminosities, \hahb\ ratio, and ultraviolet spectral slope in nearby
starburst galaxies (Charlot \& Fall 2000). Finally, we consider 
different star formation histories (as before, we assume that stars 
and gas have the same metallicity, i.e., $Z={\rm \zav}$).  We take 
the star formation rate $\psi(t)$ to be either constant or 
exponentially declining. In each case, we compute models at ages 
ranging from $10^7\,$yr to $10^{10}\,$yr. For simplicity, we adopt
in all models a Salpeter IMF truncated at 0.1 and 100~$M_\odot$ (the
line efficiency factors in Table~\ref{fittable} below would be a factor
of about 3.4 times smaller for a Scalo 1986 IMF with the same cutoff
masses).

In Table~\ref{partable}, we report the parameters of the models
found to reproduce the observations of nearby star-forming
galaxies. Most but not all combinations of these parameters are
consistent with the observations. In particular, the observed 
emission-line ratios of nearby galaxies (Figs.~1--3) do not favor
combinations of both either very low or very high \zav\ and \uavo.
In Fig.~9, we show the emission-line ratios of the models with 
the allowed combinations of \zav, \uavo, and \xav\ in 
Table~\ref{partable}. As in Figs.~1--3, we do not include in
Fig.~9 the absorption by dust in the neutral ISM and the
stellar \ha\ and \hb\ absorptions. Since age has a negligible
effect in such diagrams (especially for $t\ge10^7\,$yr; see
Figs.~1{\it a}--3{\it a}), we plot only models with constant 
$\psi (t)$ at the standard age $t= 3\times10^8 \,$yr 
(eq.~[\ref{standard}]). Fig.~9 shows that the models in 
Table~\ref{partable} account well for the typical properties, 
scatter, and trends seen in the observations. The models do not 
reach the extreme \niisii\ ratios of some \hii\ and starburst 
galaxies, suggesting that our simple prescription for secondary 
nitrogen production may break down at very low and very high \zav\
(\S2.3; we note that the largest observed \niisii\ ratios in 
Fig.~9{\it c} could result from a decrease in the S/O 
abundance ratio at high metallicity; see D\'{\i}az et al. 1991). 
We have checked that the predictions of our model at these extreme 
effective metallicities do not affect our conclusions below. The
models with various $\hat{\tau}_V$ and $\mu$ in Table~\ref{partable} 
are also consistent, by construction, with the observed ratios
of far-infrared to ultraviolet luminosities, \hahb\ ratios, and
ultraviolet spectral slopes of nearby starburst galaxies in 
Figs.~2--4 of Charlot \& Fall (2000; see \S2.4 above). Therefore,
the models in Table~\ref{partable} enable us to interpret the 
observed spectral properties of nearby star-forming galaxies in
terms of stars, gas, and dust parameters.

\begin{table}
 \caption{Parameters of the models reproducing the observed 
integrated spectral properties of nearby star-forming galaxies.}
 \label{partable}
 \begin{tabular}{@{}ll}
 $\log({\rm \uavo})$ & $-3.0$, $-2.5$, $-2.0$, $-1.5$ \\
 ${\rm \zav}/Z_\odot$ & 0.2, 0.5, 0.75, 1.0, 1.5, 2.0 \\
 ${\rm \xav}$ & 0.1, 0.3, 0.5 \\
 $\hat{\tau}_V$ & 0.01, 0.3, 0.5, 1.0, 1.5, 2.0 \\
 $\mu^{-1}$ & 1, 3, 5 \\
 $\psi(t)$ & constant, exponentially declining \\
 $t$/yr & $10^7$--$10^{10}$ (112 uneven steps) \\
\end{tabular}

\medskip
Two timescales are adopted for the exponentially declining star formation
rate $\psi(t)$: 0.1 and 6.0~Gyr. All models have $\tilde{n}_{\rm H}=
30~$cm$^{-3}$ and a Salpeter IMF truncated at 0.1 and 100~$M_\odot$.
Most but not all combinations of the above parameters are allowed by
the observations (see text for detail).
\end{table}

\begin{figure}
%\vspace{250pt}
\epsfxsize=8.6cm
\hfill{\epsfbox{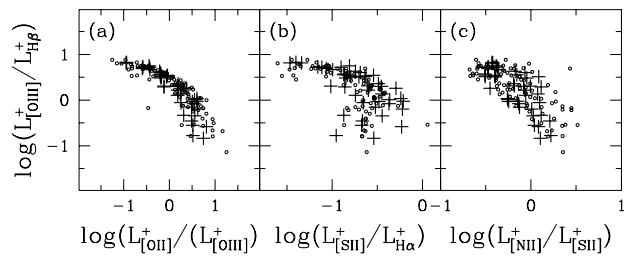}}
 \caption{Emission-line ratios of the models in Table~\ref{partable}.
The absorption by dust in the neutral ISM and the stellar \ha\ and 
\hb\ absorptions are not included. Only models with constant 
$\psi(t)$ are shown at the standard age $t=3\times10^8\,$yr
(eq.~[\ref{standard}]; {\it crosses}). The data points in ({\it a}),
({\it b}), and ({\it c}) are the same galaxies as in Figs.~1, 2, and
3, respectively (all types plotted as {\it circles} here).}
 \end{figure}

A striking property of these models is that they span vast 
ranges of \ha\ and \oii\ efficiency factors, ${\rm \etaha}
=0.05$--$2.8 \times10^{15} \,$erg~g$^{-1}$ and ${\rm \etaoii}
= 0.02$--$3.6 \times10^{15}\,$erg~g$^{-1}$ (for \wha\ and 
\woii\ both larger than 5~{\AA}). Only parts of these ranges
arise from variations in $\hat{\tau}_V$ and $\mu$ in 
Table~\ref{partable}. For $\hat{\tau}_V =0$, changes in
the ionized-gas parameters alone (\zav, \uavo, and \xav)
account for factors of 3.5 and 14 spreads in \etaha\
and \etaoii, respectively. The range of stellar \ha\ absorption
strengths arising from variations in the star formation history
contributes an additional factor of 1.8 spread in \etaha.
These results put into perspective the approximate efficiency 
factors often adopted to estimate star formation rates in 
galaxies, e.g., ${\rm \etaha} \approx 2.0\times 10^{15}\, 
$erg~g$^{ -1}$ and ${\rm \etaoii} \approx 1.2 \times 10^{15}\,
$erg~g$^{ -1}$ (Kennicutt 1998). Our model allows us to assess
that the actual uncertainties in SFR estimates based on either
the \ha\ line or the \oii\ line alone can be as high as several
decades.

Since we know how \etaha\ and \etaoii\ are related to observed
optical-line ratios (\S3.1), we can exploit this information to
improve SFR estimates in galaxies. For the sake of generality,
we denote here by $\cal{A}$ the quantity to be estimated (e.g., \etaha,
\etaoii). Given a set of $N$ spectral features $x_k$ (e.g.,
line ratios, equivalent widths), we can always least-squares fit 
to the models in Table~\ref{partable} an expression of $\cal{A}$ of the
form
\begin{equation}
\log{\cal{A}}\approx\,a_0\,+\,\sum_{k=1}^{N}\,a_k \log x_k\,,
\label{fitting}
\end{equation}
where the $a_k$ are the fitting coefficients. It is convenient
to introduce the logarithmic difference $\Delta_m$ between the 
true and fitted $\cal{A}$ values of the $m$th model,
\begin{equation}
\Delta_m= \log{{\cal{A}}_m}\,-\,a_0\,-\,\sum_{k=1}^{N}\,a_k \log x_{k,m}\,.
\label{error}
\end{equation}
The probability distribution of the models being unknown, the 
maximum absolute value of $\Delta_m$ over all the models in
Table~\ref{partable} provides a lower limit on the accuracy of 
expression~(\ref{fitting}). By experimenting with several
different combinations of spectral features $x_k$, we can identify
the one leading to the highest accuracy and hence the best 
observational estimate of $\cal{A}$.

We have used this procedure to derive estimators of not only
the \ha\ and \oii\ efficiency factors, but also the gas-phase 
oxygen abundance and the effective absorption optical depth of
the dust in star-forming galaxies. In Table~\ref{fittable},
we present our results for several examples of available 
spectral information (cases A--G). The observations are
assumed {\it not} to be corrected for dust and stellar (H-Balmer)
absorption. In each case, we express the efficiency factor 
of the most accurate SFR tracer (\etaha, \etahanii, or \etaoii),
the gas-phase oxygen abundance (\oh), and the effective 
dust-absorption factor at 5500~{\AA} (\etauv) as functions of
the available spectral features. For reference, we also give the 
expression of the line efficiency factor for $\hat{\tau}_V=0$
(noted \oetaha, \oetahanii, or \oetaoii). 

The accuracy of an estimator in Table~\ref{fittable} depends on
its regime of application. In Fig.~10, we plot for each estimator
the range in $\Delta_m$, as defined by equation (\ref{error}), 
as a function of the estimated quantity. The potential (logarithmic)
error in an estimate of $\eta_i$, \oh, or \etauv\ -- ignoring 
observational errors -- is the maximum absolute value of $\Delta_m$
at the abscissa corresponding to that estimate (we note that, for 
${\rm \etauv}\sim1$, even small values of $|\Delta_m|$ can translate
into large errors in $\hat{\tau}_V$). The maximum absolute value of
$\Delta_m$ over the full range in the estimated quantity is the 
maximum error associated to the estimator. To avoid the undue 
influence of large tails arising from a few discrepant models in the
$|\Delta_m|$ distributions, we define the `uncertainty' pertaining to
each estimator in Fig.~10 as the 99th percentile range in $|\Delta_m|$
over the full range in the estimated quantity. We report this
uncertainty in Table~\ref{fittable} and use it below to quantify
the accuracy of an estimator (for reference, we also indicate in
parentheses in Table~\ref{fittable} the full, 100th percentile 
range in $|\Delta_m|$).

The results of Table~\ref{fittable} demonstrate the merit
of spectral information for estimating the star formation rate,
the metallicity, and the absorption by dust in a galaxy. In the
best-constrained case where the \oii, \hb, \oiii, \ha, \nii, and
\sii\ lines are all observed (case~A), we find that \etaha,
\oh, and \etauv\ can be estimated to within less than a factor
of 2 (i.e., with absolute logarithmic uncertainties less than 0.3).
This is a reduction by more than a factor of 25 of the 
uncertainty in \etaha\ relative to the case where the \ha\ line 
alone is used to estimate the SFR (see above). The reason for such
an improvement is that we have included in the estimator the \hahb,
\oiioiii, \oiiihb, and \siiha\ ratios and the \hb\ equivalent width,
which all carry important information about the parameters that 
influence \etaha\ (see \S3).  Remarkably, a similar accuracy in
\etaha\ can be achieved even if the \ha\ line is blended with the
adjacent \nii\ lines (case~B).  In this case, however, the loss of
the \niisii\ ratio makes \oh\ estimates based on the \oiiihb\ and
\oiioiii\ ratios more uncertain, especially for low \oh\ (see 
Fig.~10; we do not discuss here the \oiii$\lambda{4363}$ line, 
which is a diagnostic of low oxygen abundance but is very weak in
most galaxies; see Kobulnicky et al. 1999 and Figs.~5--7 above).
In case~C, the unavailability of \hb\ (and hence \hahb) makes the 
uncertainties in \etauv\ larger, but \etaha\ and \oh\ can still 
be determined to within a factor of 2--3 using the \oii, \oiii, 
\ha, and \nii\ lines. As above, we find that the blending of \ha\
with the adjacent \nii\ lines affects only weakly the accuracy in
\etaha\ (case~D).  In the case where the \hb\ line is observed but
the \oii\ line is not (case~E), \etaha\ and \oh\ can again be 
estimated to within a factor of 2--3 using the \hb, \oiii, \ha, 
and \nii\ lines, while the \hahb\ ratio helps constrain \etauv.

\begin{table*}
 \label{fittable}
 \centering
\begin{minipage}{\hsize}
%\begin{minipage}{160mm}
  \caption{Estimators of the \ha\ and \oii\ efficiency factors, the gas-phase
oxygen abundance, and the effective absorption optical depth of the dust in 
galaxies for different assumptions about the available spectral information
(cases~A--G).}
  \label{fittable}
  \begin{tabular}{@{}lccccccclc@{}}
    	Case
	&\multicolumn{7}{c}{Available spectral information\footnote{Observed line 
luminosities and \dbal\ (as defined by Bruzual 1983) not corrected for dust and stellar
(H-Balmer) absorption. A `yes' means that the observed rest-frame equivalent width of the
line exceeds 5~{\AA}, a `no' that the equivalent width is less than this limit, and an 
ellipsis that the line is not observed.  The lines are \oii$\lambda{3727}$, 
\oiii$\lambda{5007}$, \nii$\lambda6583$ (see however footnote $e$), and
\sii$\lambda\lambda{6717,\, 6731}$.}}
        &Diagnostics\footnote{The basis functions $x_k$ are spectral features normalized
to the same standard model as in eq.~[\ref{standard}] but including the effects of stellar
\ha\ and \hb\ absorption (for $\hat{\tau}_V=0$):	     
			      $x_1\equiv(\rm{\lhahb})/3.4$;
                              $x_2\equiv(\rm{\loiioiii})/1.5$;
                              $x_3\equiv(\rm{\loiiihb})/2.0$;
                              $x_4\equiv(\rm{\lsiiha})/0.26$;
			      $x_5\equiv{\rm{\whb}}/(25\,{\rm \AA})$;
                              $x_6\equiv(\rm{\lniisii})/0.85$;
			      $x_7\equiv(\rm{\lhaniihb})/4.5$;
			      $x_8\equiv(\rm{\lsiihanii})/0.20$;
			      $x_9\equiv(\rm{\loiiha})/0.90$;
			      $x_{10}\equiv(\rm{\lniiha})/0.22$;
			      $x_{11}\equiv{\rm{\wha}}/(170\,{\rm \AA})$;
			      $x_{12}\equiv(\rm{\loiihanii})/0.70$;
			      $x_{13}\equiv{\rm{\whanii}}/(220\,{\rm \AA})$;
			      $x_{14}\equiv(\rm{\loiiihanii})/0.45$;
			      $x_{15}\equiv(\rm{\loiihb})/3.0$;
			      $x_{16}\equiv{\rm{\woii}}/(60\,{\rm \AA})$;
			      $x_{17}\equiv{\dbal}/1.1$.}
         and estimators\footnote{Line efficiency factors \etaha, \etaoii, and
\etahanii\ in units of erg~g$^{-1}$ (eq.~[\ref{effpar}]); oxygen abundance by 
number relative to hydrogen in the ionized gas after accounting for the 
depletion onto dust grains [(O/H)$_{\rm cosmic}=6.6\times10^{-4}$]; and 
effective dust-absorption factor \etauv\ at 5500~{\AA} (eqs.~[\ref{taudef}] and
[\ref{taueff}]). The line efficiency factors \oetaha, \oetaoii, and 
\oetahanii\ correspond to the case $\hat{\tau}_V=0$.}
        &{Uncertainty}\footnote{99th percentile range in $|\Delta_m|$ 
(eq.~[\ref{error}]) obtained when applying the formula to all the models $m$
in Table~\ref{partable} (see text and Fig.~10). For reference, the full (100th
percentile) range in $|\Delta_m|$ is indicated in parentheses.}
        \\
%%%%%%%%%%%%%%%%%%%%%%%%%%%%%%%%%%%%%%%%%%%%%%%%%%%%%%%%%%%%%%%%%%%%%%%%%%%%%%%%%%%%%%%%%%%%%%%%
  & \oii\ & \dbal\ & \hb\ & \oiii\ & \ha\ & \nii\ & \sii\
        &
        & (logarithmic)
\\[10pt]
%%%%%%%%%%%%%%%%%%%%%%%%%%%%%%%%%%%%%%%%%%%%%%%%%%%%%%%%%%%%%%%%%%%%%%%%%%%%%%%%%%%%%%%%%%%%%%%%
A &   yes & yes & yes & yes & yes & yes & yes &
${\rm \etaha}=
1.85\times10^{15}\,x_1^{-2.25}\,x_2^{1.46}\,x_3^{0.90}\,x_4^{-1.14}\,x_5^{-0.15}$ &
0.26~~(0.39)\\[3pt]
%%%%%%%%%%%%%%%%%%%%%%%%%%%%%%%%%%%%%%%%%%%%%%%%%%%%%%%%%%%%%%%%%%%%%%%%%%%%%%%%%%%%%%%%%%%%%%%%
  & & & & & & & &
${\rm \oh}=5.09\times10^{-4}\,x_2^{0.17}\,x_6^{1.17}$ & 
0.24~~(0.26)\\[3pt]
%%%%%%%%%%%%%%%%%%%%%%%%%%%%%%%%%%%%%%%%%%%%%%%%%%%%%%%%%%%%%%%%%%%%%%%%%%%%%%%%%%%%%%%%%%%%%%%%
  & & & & & & & &
${\rm \etauv}= 1.20\,x_1^{-3.32}\,x_2^{1.06}\,x_3^{0.62}\,x_4^{-0.85}\,x_5^{-0.49}$ & 
0.27~~(0.46)\\[3pt]
%%%%%%%%%%%%%%%%%%%%%%%%%%%%%%%%%%%%%%%%%%%%%%%%%%%%%%%%%%%%%%%%%%%%%%%%%%%%%%%%%%%%%%%%%%%%%%%%
  & & & & & & & &
${\rm \oetaha}=
1.44\times10^{15}\,x_1^{0.75}\,x_3^{-0.17}\,x_5^{0.15}\,x_6^{-0.58}$ &
0.14~~(0.20)\\[9pt]
%%%%%%%%%%%%%%%%%%%%%%%%%%%%%%%%%%%%%%%%%%%%%%%%%%%%%%%%%%%%%%%%%%%%%%%%%%%%%%%%%%%%%%%%%%%%%%%%
B & yes & yes & yes & yes &\multicolumn{2}{c}{blended\footnote{\hanii$\lambda\lambda6548,\,
6583$ blend only (${\rm \whanii}>5\,${\AA}).}}& yes &
${\rm \etahanii}=
2.29\times10^{15}\,x_2^{1.80}\,x_3^{1.04}\,x_7^{-1.40}\,x_8^{-1.35}$ &
0.33~~(0.39)\\[3pt]
%%%%%%%%%%%%%%%%%%%%%%%%%%%%%%%%%%%%%%%%%%%%%%%%%%%%%%%%%%%%%%%%%%%%%%%%%%%%%%%%%%%%%%%%%%%%%%%%
  & & & & & & & &
${\rm \oh}=4.21\times10^{-4}\,x_2^{0.51}\,x_3^{-0.24}\,x_8^{-0.42}$ \hfill (${\rm \loiioiii}<0.8$) & 
0.52~~(0.58)\\[3pt]
%%%%%%%%%%%%%%%%%%%%%%%%%%%%%%%%%%%%%%%%%%%%%%%%%%%%%%%%%%%%%%%%%%%%%%%%%%%%%%%%%%%%%%%%%%%%%%%%
  & & & & & & & &
${\rm \oh}=4.58\times10^{-4}\,x_3^{-0.37}x_8^{-0.54}$ \hfill (${\rm \loiioiii}\ge0.8$) & 
0.21~~(0.26)\\[3pt]
%%%%%%%%%%%%%%%%%%%%%%%%%%%%%%%%%%%%%%%%%%%%%%%%%%%%%%%%%%%%%%%%%%%%%%%%%%%%%%%%%%%%%%%%%%%%%%%%
  & & & & & & & &
${\rm \etauv}=0.90\,x_2^{2.01}\,x_3^{1.21}\,x_7^{-1.41}\,x_8^{-1.70}$ &
0.32~~(0.38)\\[3pt]
%%%%%%%%%%%%%%%%%%%%%%%%%%%%%%%%%%%%%%%%%%%%%%%%%%%%%%%%%%%%%%%%%%%%%%%%%%%%%%%%%%%%%%%%%%%%%%%%
  & & & & & & & &
${\rm \oetahanii}=
2.31\times10^{15}\,x_2^{0.28}\,x_3^{0.12}\,x_7^{-0.36}\,x_8^{-0.06}$ &
0.25~~(0.26)\\[9pt]
%%%%%%%%%%%%%%%%%%%%%%%%%%%%%%%%%%%%%%%%%%%%%%%%%%%%%%%%%%%%%%%%%%%%%%%%%%%%%%%%%%%%%%%%%%%%%%%%

C & yes & yes & no & yes & yes & yes & \ldots &
${\rm \etaha}=
1.36\times10^{15}\,x_2^{0.38}\,x_9^{0.80}\,x_{10}^{-0.40}\,x_{11}^{0.33}$ &
0.41~~(0.52)\\[3pt]
%%%%%%%%%%%%%%%%%%%%%%%%%%%%%%%%%%%%%%%%%%%%%%%%%%%%%%%%%%%%%%%%%%%%%%%%%%%%%%%%%%%%%%%%%%%%%%%%
  & & & & & & & &
${\rm \oh}=3.98\times10^{-4}\,x_2^{-0.21}\,x_{10}^{0.39}$ &
0.31~~(0.40)\\[3pt]
%%%%%%%%%%%%%%%%%%%%%%%%%%%%%%%%%%%%%%%%%%%%%%%%%%%%%%%%%%%%%%%%%%%%%%%%%%%%%%%%%%%%%%%%%%%%%%%%
  & & & & & & & &
${\rm \etauv} = 0.64\,x_2^{-0.12}\,x_9^{0.62}\,x_{11}^{0.20}$ &
0.49~~(0.54)\\[3pt]
%%%%%%%%%%%%%%%%%%%%%%%%%%%%%%%%%%%%%%%%%%%%%%%%%%%%%%%%%%%%%%%%%%%%%%%%%%%%%%%%%%%%%%%%%%%%%%%%
  & & & & & & & &
${\rm \oetaha}=
1.69\times10^{15}\,x_2^{0.64}\,x_9^{-0.10}\,x_{10}^{-0.45}\,x_{11}^{0.09}$ &
0.12~~(0.15)\\[9pt]
%%%%%%%%%%%%%%%%%%%%%%%%%%%%%%%%%%%%%%%%%%%%%%%%%%%%%%%%%%%%%%%%%%%%%%%%%%%%%%%%%%%%%%%%%%%%%%%%

D & yes & yes & no & yes &\multicolumn{2}{c}{blended$^e$} & \ldots &
${\rm \etahanii}=
2.99\times10^{15}\,x_2^{-0.12}\,x_{12}^{0.86}\,x_{13}^{0.68}$ &
0.52~~(0.66)\\[3pt]
%%%%%%%%%%%%%%%%%%%%%%%%%%%%%%%%%%%%%%%%%%%%%%%%%%%%%%%%%%%%%%%%%%%%%%%%%%%%%%%%%%%%%%%%%%%%%%%%
  & & & & & & & &
${\rm \oh}=4.32\times10^{-4}\,x_2^{0.13}\,x_{14}^{-0.43}$ \hfill (${\rm \loiioiii}<0.8$) &
0.52~~(0.56)\\[3pt]
%%%%%%%%%%%%%%%%%%%%%%%%%%%%%%%%%%%%%%%%%%%%%%%%%%%%%%%%%%%%%%%%%%%%%%%%%%%%%%%%%%%%%%%%%%%%%%%%
  & & & & & & & &
${\rm \oh}=3.63\times10^{-4}\,x_2^{-0.57}x_{14}^{-0.72}$ \hfill (${\rm \loiioiii}\ge0.8$) &
0.27~~(0.41)\\[3pt]
%%%%%%%%%%%%%%%%%%%%%%%%%%%%%%%%%%%%%%%%%%%%%%%%%%%%%%%%%%%%%%%%%%%%%%%%%%%%%%%%%%%%%%%%%%%%%%%%
  & & & & & & & &
${\rm \etauv} = 0.82\,x_2^{-0.17}\,x_{12}^{0.61}\,x_{13}^{0.35}$ &
0.50~~(0.59)\\[3pt]
%%%%%%%%%%%%%%%%%%%%%%%%%%%%%%%%%%%%%%%%%%%%%%%%%%%%%%%%%%%%%%%%%%%%%%%%%%%%%%%%%%%%%%%%%%%%%%%%
  & & & & & & & &
${\rm \oetahanii}=
2.09\times10^{15}\,x_{13}^{0.14}\,x_{14}^{-0.13}$ &
0.25~~(0.30)\\[9pt]
%%%%%%%%%%%%%%%%%%%%%%%%%%%%%%%%%%%%%%%%%%%%%%%%%%%%%%%%%%%%%%%%%%%%%%%%%%%%%%%%%%%%%%%%%%%%%%%%

E & \ldots & \ldots & yes & yes & yes & yes & \ldots &
${\rm \etaha}=
1.71\times10^{15}\,x_1^{-2.96}\,x_3^{-0.12}\,x_5^{-0.17}\,x_{10}^{-0.18}\,x_{11}^{-0.38}$ &
0.45~~(0.57)\\[3pt]
%%%%%%%%%%%%%%%%%%%%%%%%%%%%%%%%%%%%%%%%%%%%%%%%%%%%%%%%%%%%%%%%%%%%%%%%%%%%%%%%%%%%%%%%%%%%%%%%
  & & & & & & & &
${\rm \oh}=4.15\times10^{-4}\,x_3^{-0.29}\,x_{10}^{0.40}$ \hfill (${\rm \loiiihb}\le2.5$) &
0.32~~(0.33)\\[3pt]
%%%%%%%%%%%%%%%%%%%%%%%%%%%%%%%%%%%%%%%%%%%%%%%%%%%%%%%%%%%%%%%%%%%%%%%%%%%%%%%%%%%%%%%%%%%%%%%%
  & & & & & & & &
${\rm \oh}=4.15\times10^{-4}\,x_3^{-0.29}\,x_{10}^{0.40}$ \hfill (${\rm \loiiihb}>2.5$) &
0.40~~(0.40)\\[3pt]
%%%%%%%%%%%%%%%%%%%%%%%%%%%%%%%%%%%%%%%%%%%%%%%%%%%%%%%%%%%%%%%%%%%%%%%%%%%%%%%%%%%%%%%%%%%%%%%%
  & & & & & & & &
${\rm \etauv} =1.15\,x_1^{-3.97}\,x_3^{-0.01}\,x_5^{-1.21}\,x_{10}^{0.01}\,x_{11}^{0.52}$ &
0.35~~(0.49)\\[3pt]
%%%%%%%%%%%%%%%%%%%%%%%%%%%%%%%%%%%%%%%%%%%%%%%%%%%%%%%%%%%%%%%%%%%%%%%%%%%%%%%%%%%%%%%%%%%%%%%%
  & & & & & & & &
${\rm \oetaha}=
1.34\times10^{15}\,x_1^{2.18}\,x_3^{-0.12}\,x_5^{2.44}\,x_{10}^{-0.17}\,x_{11}^{-2.13}$ &
0.21~~(0.35)\\[9pt]
%%%%%%%%%%%%%%%%%%%%%%%%%%%%%%%%%%%%%%%%%%%%%%%%%%%%%%%%%%%%%%%%%%%%%%%%%%%%%%%%%%%%%%%%%%%%%%%%

F & yes & yes & yes & yes & \ldots & \ldots & \ldots &
${\rm \etaoii}=
6.92\times10^{14}\,x_{2}^{-0.02}\,x_{5}^{0.41}\,x_{15}^{1.52}$ &
0.70~~(0.99)\\[3pt]
%%%%%%%%%%%%%%%%%%%%%%%%%%%%%%%%%%%%%%%%%%%%%%%%%%%%%%%%%%%%%%%%%%%%%%%%%%%%%%%%%%%%%%%%%%%%%%%%
  & & & & & & & &
${\rm \oh}=3.78\times10^{-4}\,x_2^{0.17}\,x_3^{-0.44}$ \hfill (${\rm \loiioiii}<0.8$) &
0.51~~(0.57)\\[3pt]
%%%%%%%%%%%%%%%%%%%%%%%%%%%%%%%%%%%%%%%%%%%%%%%%%%%%%%%%%%%%%%%%%%%%%%%%%%%%%%%%%%%%%%%%%%%%%%%%
  & & & & & & & &
${\rm \oh}=3.96\times10^{-4}\,x_3^{-0.46}$ \hfill (${\rm \loiioiii}\ge0.8$) &
0.31~~(0.31)\\[3pt]
%%%%%%%%%%%%%%%%%%%%%%%%%%%%%%%%%%%%%%%%%%%%%%%%%%%%%%%%%%%%%%%%%%%%%%%%%%%%%%%%%%%%%%%%%%%%%%%%
  & & & & & & & &
${\rm \etauv} =0.47\,x_2^{0.05}\,x_{5}^{0.01}$ &
0.55~~(0.56)\\[3pt]
%%%%%%%%%%%%%%%%%%%%%%%%%%%%%%%%%%%%%%%%%%%%%%%%%%%%%%%%%%%%%%%%%%%%%%%%%%%%%%%%%%%%%%%%%%%%%%%%
  & & & & & & & &
${\rm \oetaoii}=
1.43\times10^{15}\,x_{2}^{-1.01}\,x_{3}^{-1.03}\,x_{5}^{-1.21}\,x_{16}^{1.99}$ &
0.20~~(0.40)\\[9pt]
%%%%%%%%%%%%%%%%%%%%%%%%%%%%%%%%%%%%%%%%%%%%%%%%%%%%%%%%%%%%%%%%%%%%%%%%%%%%%%%%%%%%%%%%%%%%%%%%

G & yes & yes & \ldots & \ldots & \ldots & \ldots & \ldots &
${\rm \etaoii}=
4.44\times10^{14}\,x_{16}^{0.73}\,x_{17}^{1.87}$ &
0.94~~(1.47)\\[3pt]
%%%%%%%%%%%%%%%%%%%%%%%%%%%%%%%%%%%%%%%%%%%%%%%%%%%%%%%%%%%%%%%%%%%%%%%%%%%%%%%%%%%%%%%%%%%%%%%%
  & & & & & & & &
${\rm \oh}=3.34\times10^{-4}\,x_{17}^{-0.23}$ &
0.63~~(0.63)\\[3pt]
%%%%%%%%%%%%%%%%%%%%%%%%%%%%%%%%%%%%%%%%%%%%%%%%%%%%%%%%%%%%%%%%%%%%%%%%%%%%%%%%%%%%%%%%%%%%%%%%
  & & & & & & & &
${\rm \etauv} =0.47\,x_{16}^{0.10}\,x_{17}^{-0.86}$ &
0.56~~(0.61)\\[3pt]
%%%%%%%%%%%%%%%%%%%%%%%%%%%%%%%%%%%%%%%%%%%%%%%%%%%%%%%%%%%%%%%%%%%%%%%%%%%%%%%%%%%%%%%%%%%%%%%%
  & & & & & & & &
${\rm \oetaoii}=
1.21\times10^{15}\,x_{16}^{0.89}\,x_{17}^{3.78}$ &
0.39~~(0.96)\\[9pt]
%%%%%%%%%%%%%%%%%%%%%%%%%%%%%%%%%%%%%%%%%%%%%%%%%%%%%%%%%%%%%%%%%%%%%%%%%%%%%%%%%%%%%%%%%%%%%%%%

\end{tabular}
\end{minipage}
\end{table*}

\begin{figure}
%\vspace{250pt}
\epsfxsize=9cm
\hfill{\epsfbox{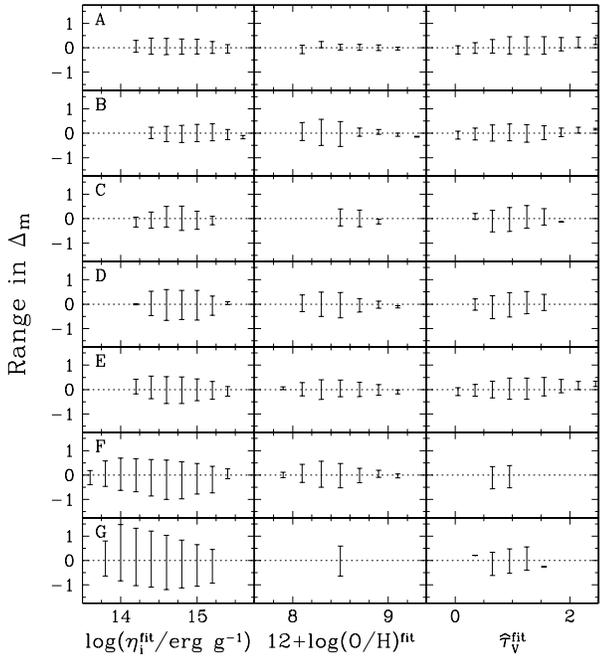}}
 \caption{Range in the logarithmic error $\Delta_m$ (as defined
by eq.~[\ref{error}]) obtained when applying the formulae in 
Table~\ref{fittable} to the models in Table~\ref{partable}, plotted 
against estimated line efficiency factor, gas-phase oxygen abundance,
and effective absorption optical depth of the dust. Each panel row
corresponds to a different case in Table~\ref{fittable}, as indicated.}
\end{figure}

A critical situation arises when the \ha\ line is not available
(cases~F and G in Table~\ref{fittable}). We find that, in this
case, the star formation rate cannot be determined reliably from
the \oii, \hb, and \oiii\ lines and the 4000~{\AA} discontinuity.
Using all these features, we did not succeed in estimating 
\etaoii\ (or \etahb\ or \etaoiii) to within less than a factor of 
5 (case~F). In the case where only \oii\ and \dbal\ are available 
(case~G), the uncertainty in \etaoii\ reaches a factor of almost 10,
and O/H is unconstrained (Fig.~10). The major source of uncertainties
in \etaoii\ estimates is the lack of a strong dust discriminant
in the wavelength interval between \oii\ and \oiii.  The
uncertainties in \oetaoii\ in cases~F and G demonstrate that,
without dust, the \oii\ efficiency factor can be constrained
to within a factor of only 2--3 based simply on the \oii\ 
equivalent width and the 4000~{\AA} discontinuity. As a further
check, we reestimated \etaoii\ in case~F after reducing the 
ranges in zero-age effective ionization parameter and effective
metallicity of the models in Table~\ref{fittable} to $0.5
\,Z_\odot\le {\rm \zav}\le Z_\odot$ and $-3.0\le \log({\rm 
\uavo})\le -2.0$. The resulting uncertainty was 0.65, comparable
to the value of 0.70 in Table~\ref{fittable}. We note in Fig.~10
that \etaoii\ estimates are slightly more accurate for ${\rm 
\etaoii} \la10^{14} \,$erg~g$^{-1}$, corresponding to galaxies
with high effective metallicity and high dust content, and 
${\rm \etaoii}\ga10^{15}\, $erg~g$^{-1}$, corresponding to 
galaxies with low effective metallicity and low dust content.

Independent constraints on the absorption by dust in a galaxy
are not expected to improve significantly SFR estimates based 
on the \oii\ luminosity. In nearby starburst galaxies, the 
correlation between ratio of far-infrared to ultraviolet 
luminosities and ultraviolet spectral slope sets tight constraints
on the effective absorption optical depth of the dust at ultraviolet
wavelengths (e.g., Meurer, Heckman \& Calzetti 1999; Charlot \& Fall
2000). However, the different attenuation of line and continuum 
photons in these galaxies (accounted for by values of $\mu^{-1}$
greater than unity in Table~\ref{partable}; see \S2.4) make 
corrections of the \oii\ luminosity uncertain. This is all the
more critical in that the \oii\ equivalent width is needed to 
estimate \oetaoii\ (Table~\ref{fittable}). We therefore conclude
that, without the \ha\ line, the SFR is difficult to estimate 
from the optical emission of a galaxy.

It is worth recalling that the estimators in Table~\ref{fittable}
neglect the potential contribution by active galactic nuclei to
the ionizing radiation in galaxies. The presence of an AGN in a
galaxy is usually revealed by readily identifiable signatures.
These include larger \oiiihb\ and \siiha\ ratios than in the
galaxies in Fig.~2 and strong emission lines of highly-ionized
species, such as \nev$\lambda3426$ and \fevii$\lambda6087$, 
which are generally not produced by the softer spectra of stellar
populations (e.g., Osterbrock 1989). Furthermore, the broad
emission lines of most AGNs have velocity widths (FWHMs of 
several thousand km$\,{\rm s}^{-1}$) that are several times 
larger than those expected from the virial motions within 
galaxies (FWHMs less than about 1000 km$\,{\rm s}^{-1}$, 
corresponding to line-of-sight velocity dispersions less than 
about 400 km$\,{\rm s}^{-1}$). Hence, the contamination of the 
nebular emission from a galaxy by an AGN can usually be identified.

\section{Conclusions}

We have developed a model for computing consistently the line and 
continuum emission from galaxies, based on an idealized description
of the \hii\ regions and the diffuse gas ionized by single stellar
generations in a galaxy. We have calibrated the nebular properties
of this model using the observed \oiiihb, \oiioiii, \siiha, and 
\niisii\ ratios of a representative sample of nearby spiral and 
irregular, starburst, and \hii\ galaxies. To compute whole (line
plus continuum) spectral energy distributions, we have included 
the absorption by dust in the neutral ISM using a recent simple
prescription, which is consistent with observations of nearby
starburst galaxies. The model succeeds in reproducing quantitatively
the optical spectra of nearby galaxies of various types. The prime
parameters in our model, including the star formation history, the
zero-age effective ionization parameter, the effective gas metallicity,
the effective dust-to-heavy element ratio in the ionized gas, and 
the effective optical depth of the dust in the neutral ISM (and the
fraction of this contributed by the ambient ISM) each have a specific
influence on the integrated line and continuum properties of galaxies.
Optical spectral fits enable us, in turn, to constrain simultaneously
the star formation history, the metallicity, and the absorption by
dust in galaxies.

We have used this model to derive new estimators of the \ha\ and
\oii\ efficiency factors (\etaha\ and \etaoii), the gas-phase
oxygen abundance, and the effective absorption optical depth of
the dust in star-forming galaxies. The observed integrated spectral
properties of nearby galaxies are compatible with ranges of at 
least a decade in \etaha\ and \etaoii. The allowed combinations of
ionized-gas parameters alone imply factors of 3.5 and 14 spreads in 
\etaha\ and \etaoii, respectively. We find that, by exploiting the 
correlations between \etaha\ and \etaoii\ and various spectral features,
we can reduce considerably the uncertainties affecting SFR estimates.
For example, with the help of other lines such as \oii, \hb, \oiii\,
\nii, or \sii, the uncertainties in SFR estimates based on \ha\ can
be reduced to a factor of only 2--3. A similar accuracy can be achieved
even if the \ha\ line is blended with the adjacent \nii\ lines. 
Without \ha, however, the SFR, the gas-phase oxygen abundance, and
the effective absorption optical depth of the dust are difficult to
estimate from the \oii, \hb, and \oiii\ lines. The reason for this 
is that the absorption by dust in the neutral ISM and the ionized-gas
parameters are then difficult to constrain independently. This suggests 
that, while insufficient by itself, the \ha\ line is essential for
estimating the star formation rate from the optical emission of a galaxy.

The estimators listed in Table~\ref{fittable} allow one to derive
more refined constraints on the star formation rate, the gas-phase
oxygen abundance, and the effective absorption optical depth of the
dust in galaxies than was previously possible from optical spectra 
alone. By design, these estimators can be applied directly to
observations uncorrected for dust and stellar (H-Balmer) absorption. 
We can also derive, on demand, the estimators best suited to 
observational selection criteria other than those exemplified in 
Table~\ref{fittable}.  Finally, as shown in \S3.2, our full model
for the line and continuum emission allows detailed interpretations
of whole spectral energy distributions of galaxies in terms of stars,
gas, and dust parameters.

\section*{Acknowledgments}

We thank M.~Fall, G.~Ferland, R.~Kennicutt, and L.~Tresse
for valuable discussions. S.C. appreciates the generous 
hospitality of the MPIA-Garching, and M.L. acknowledges 
support by the European Commission under TMR grant 
no.~ERBFMBI-CT97-2804. This research was supported
in part by the National Science Foundation through grant 
no.~PHY94-07194 to the Institute for Theoretical Physics.

\bsp

\label{lastpage}

\end{document}